\documentclass[journal=mamobx,manuscript=article]{achemso}
\usepackage{graphicx}
\usepackage[update,prepend]{epstopdf}
\usepackage{natmove}
\usepackage{booktabs}

\title{Interaction-Dependent Molecular Probing of Heterogeneous Glassy Dynamics in Polymer Melts}

\author{Suyeon Kim}
\affiliation{Department of Chemical Education, Jeju National University, Jeju 63243, Republic of Korea}

\author{Taejin Kwon}
\email{tjkwon@jejunu.ac.kr}
\affiliation{Department of Chemistry and Cosmetics, Jeju National University, Jeju 63243, Republic of Korea}
\begin{document}

\begin{abstract}
Single-molecule probes are widely used to characterize dynamic heterogeneity in glass-forming liquids, but their measured dynamics can depend on how the probe couples to its host environment. Using molecular dynamics simulations of dilute dimer probes embedded in a supercooled polymer melt, we examine how probe--host affinity influences the correspondence between probe rotational relaxation and host dynamics. Increasing the probe--host interaction strength systematically modifies the relative temperature dependence of probe and host relaxation while leaving the intrinsic host dynamics essentially unperturbed. Lower-affinity probes exhibit stronger non-Gaussian translational motion and larger displacement fluctuations, whereas higher-affinity probes remain more localized and are statistically less exposed to the most dynamically active host environments. Comparison with wavevector-dependent host structural relaxation further identifies an effective matching wavevector at which the host structural relaxation exhibits the temperature dependence most similar to that of the probe, providing a phenomenological measure of the probe--host dynamical correspondence. Together with real-space analyses of the local environments sampled by the probe, these results show that probe--host affinity systematically influences the quantitative relationship between molecular-probe relaxation and heterogeneous host dynamics.
\end{abstract}
\maketitle

\section{Introduction}

As polymer melts cool toward their glass transition temperature ($T_g$), their relaxation times and viscosity increase dramatically with a strongly non-Arrhenius temperature dependence\cite{williams1955,plazek1980,binder2003,Bormuth2010,Li2015,schmidtke2015,Hung2019,xu2022,yuan2025}. The underlying mechanism is widely associated with dynamic heterogeneity, which is characterized by the coexistence of spatially and temporally varying local relaxation rates\cite{bennemann1999,ediger2000,qiu2003,Stein2008,lee2009a,Cho2017,manz2018,Godey2019,xie2020,chu2021,goto2023}. Single-molecule fluorescence experiments have been widely used to investigate these heterogeneous microscopic dynamics by tracking the orientational dynamics of embedded molecular probes\cite{blackburn1996,deschenes2001,woll2009,lee2009,kulzer2010,paeng2011a,Kaufman2013,Leone2013,Paeng2014,hebert2017,huang2020,kim2023,Yadav2026}. In this approach, the interpretation of probe dynamics relies on understanding how the probe dynamics are related to the intrinsic dynamics of the host matrix.

The dynamics reported by molecular probes depends on which spatial and temporal scales of glassy relaxation they sample\cite{adhikari2007,Mackowiak2011,flier2011,paeng2011,Kuhnhold2014,Paeng2015,paeng2016,Roberts2018,manz2019,Mutneja2021,mandel2022,mandel2022a,mahato2026}. These scales range from localized sub-cage fluctuations to cooperative motions characterized by a growing dynamic correlation length ($\xi_D$) upon supercooling. Because these dynamical length scales coexist, the spatial region sampled by a probe can strongly affect the dynamics it reports. A probe can therefore reflect localized fluctuations, cooperative structural relaxation, or an average over heterogeneous environments depending on how it couples to the host matrix.

Previous studies have established that probe size and geometry can play important roles in determining how molecular probes sample heterogeneous host dynamics \cite{Liu2015,Zhang2016,Choi2019,Popova2020,Kim2022,Lin2024}. Larger probes can average over multiple local environments, leading to nearly exponential relaxation decays that reduce sensitivity to local heterogeneous dynamics \cite{adhikari2007,Mackowiak2011a,Li2023}. Smaller mobile probes can exhibit enhanced translational mobility or hopping-like motion relative to the structural relaxation of the surrounding matrix \cite{Liu2015}. Related simulations of small penetrants in polymer matrices further show that transport near the glass transition is governed by activated hopping-like rearrangements and depends strongly on penetrant size and penetrant--polymer coupling \cite{Layding2026}. More generally, heterogeneous glassy dynamics can give rise to decoupling between translational and rotational relaxation even in the absence of a probe--host size mismatch \cite{Swallen2003}. Covalent tethering can also modify how a probe samples its surroundings by restricting translation while retaining rotational sensitivity to polymer segmental dynamics and providing spatial selectivity in heterogeneous or interfacial environments \cite{Vallee2010b,Choi2019}.

Physicochemical interactions provide another route by which the local environments sampled by a probe can be altered. Probe--host affinity can be tuned through molecular features such as polarity, hydrogen-bonding capability, and dispersion interactions\cite{Diachun1994,Hall1999,Carmer2011,Zhang2017,Zhu2023,Li2023}. Strong intermolecular interactions have often been interpreted as increasing the effective size or coupling strength of the probe. Such interactions may also modify the local dynamical environments sampled by the probe and thereby influence how probe dynamics reflect the heterogeneous dynamics of the host \cite{Mackowiak2011,Kaufman2013}. These findings motivate a more direct examination of how nonbonded probe--host interactions, independent of changes in probe geometry, influence the local dynamical environments sampled by the probe and its quantitative correspondence with host relaxation.

In this work, we employ dilute probe dimers embedded in a supercooled model polymer melt, following the minimal anisotropic probe framework used in previous molecular dynamics studies of polymer melts near the glass transition\cite{Vallee2010,Vallee2010b}. The dimer serves as a minimal anisotropic model rather than a detailed representation of a specific experimental fluorescent probe, providing a well-defined orientational degree of freedom that allows the effect of probe--host affinity to be examined independently. We characterize the rotational relaxation of the probe and compare its temperature dependence directly with that of the host polymer. We further combine real-space analysis of the dynamical environments sampled by the probe with wavevector-dependent host structural relaxation and dynamic susceptibility. Through these analyses, we examine how probe--host affinity modifies the quantitative correspondence between probe and host dynamics and identify an effective wavevector at which the host relaxation exhibits a temperature dependence most similar to that of the probe. Our results provide a microscopic picture of interaction-dependent sampling of heterogeneous host dynamics and a phenomenological framework for relating probe relaxation to the wavevector-dependent dynamics of the polymer matrix.

\section{Methods and Models}

We employ a coarse-grained, generic bead--spring model to simulate the polymer melts and probe molecules\cite{Vallee2010,Vallee2010b}. The polymers are modeled as linear chains of \(N = 32\) monomers, each with mass \(m\). This chain length allows us to focus on segmental and local structural relaxation without introducing long-time entanglement-dominated dynamics. The model has been extensively used in previous studies of glass-forming liquids\cite{Riggleman2006,xu2016,Cho2017,Hsu2019,Kwon2020,xu2022,Layding2026}. Our system consists of 400 polymer chains and 8 probe molecules in a cubic simulation box with periodic boundary conditions applied in all directions. The probe molecules are modeled as two-bead dimers that provide a minimal anisotropic probe with a well-defined orientational degree of freedom (Figure~\ref{fig1}). To minimize perturbation of the intrinsic dynamics of the host polymer matrix, the probe concentration is kept sufficiently dilute. For the primary probe model, this corresponds to a volume fraction of $\phi \approx 7 \times 10^{-4}$. Additionally, a pure polymer melt without probe molecules is simulated to provide a baseline for comparison. All simulations are performed using the Large-scale Atomic/Molecular Massively Parallel Simulator (LAMMPS) in standard Lennard-Jones (LJ) reduced units\cite{Thompson2022}, where the energy unit is $\varepsilon$, the length unit is $\sigma$, the mass unit is $m$, and the corresponding time unit is $\tau = \sigma\sqrt{m/\varepsilon}$. The equations of motion are integrated with a time step of $\Delta t_{MD} = 0.005\tau$. Simulations are carried out in the isothermal--isobaric (NPT) ensemble at zero external pressure ($P=0$) using a Nos\'e--Hoover thermostat and barostat.

The nonbonded interaction ($U_{\mathrm{nonbond}}(r)$) between any two particles is described by a truncated and shifted LJ potential:
\begin{equation}
U_{\mathrm{nonbond}}(r) = 4\varepsilon_{ab} \left[ \left(\frac{\sigma_{ab}}{r}\right)^{12} - \left(\frac{\sigma_{ab}}{r}\right)^6 \right] - \varepsilon_{c}, \quad r < r_c.
\end{equation}
Here, $r$ denotes the distance between two particles, and the subscripts $a, b \in \{p, d\}$ represent polymer and probe beads, respectively. The cutoff distance is set to $r_c = 2.5\sigma_{ab}$, and $\varepsilon_c$ is the potential energy evaluated at $r_c$ to shift the potential to zero at the cutoff. The polymer--polymer and dimer--dimer interaction strengths are fixed at \(\varepsilon_{pp} = \varepsilon_{dd} = 1\varepsilon\). For the primary probe model, \(\sigma_{pp}=\sigma_{dd}=1\sigma\) and \(\sigma_{pd}=1\sigma\). To examine probe-size effects, additional dimer sizes of \(\sigma_{dd}=s\sigma\) with \(s=1.5\) and \(2\) were considered, with \(\sigma_{pd}=(\sigma_{pp}+\sigma_{dd})/2\) according to the Lorentz mixing rule. To systematically vary the effective probe--host affinity, the probe--polymer interaction strength (\(\varepsilon_{pd}\)) is varied from \(1\varepsilon\) to \(2\varepsilon\). This range spans from a probe--host interaction strength matched to the polymer--polymer interaction strength to twice that value, while the system-averaged host relaxation remains essentially unchanged. Within this coarse-grained model, \(\varepsilon_{pd}\) is interpreted as an effective isotropic probe--host affinity rather than as a specific microscopic interaction energy. 

The bonded interaction ($U_{bond}(r)$) between adjacent particles in a polymer or a dimer is modeled by the standard FENE potential with a repulsive LJ contribution:
\begin{equation}
U_{bond}(r) =
-\frac{1}{2} K R_0^2
\ln \left[ 1 - \left(\frac{r}{R_0}\right)^2 \right]
+ 4\varepsilon
\left[
\left(\frac{\sigma}{r}\right)^{12}
-
\left(\frac{\sigma}{r}\right)^6
\right]
+ \varepsilon .
\end{equation}
Here, the spring constant is $K = 30\varepsilon/\sigma^2$, and the maximum allowable bond length is $R_0 = 1.5\sigma$. The repulsive LJ contribution is truncated at $r = 2^{1/6}\sigma$ and shifted to zero at the cutoff, whereas the FENE term is defined for $r < R_0$. For each combination of temperature and $\varepsilon_{pd}$, 3 to 6 independent trajectories are generated.

Simulations were performed over a wide temperature range, from the high-temperature liquid ($T = 2.0$) down to the deeply supercooled regime. The low-temperature relaxation analyses presented here focus on $T = 0.42$--$0.50$, which is the range over which the probe and host relaxation times are directly sampled; for the most strongly attractive probes ($\varepsilon_{pd} = 1.8\varepsilon$ and $2\varepsilon$), the lowest accessible temperature is $T = 0.425$. At each temperature and probe--polymer interaction strength, the system was equilibrated until the root-mean-square displacement of the polymer chain centers of mass became comparable to the polymer radius of gyration, $R_g$. Production trajectories were then generated for durations of approximately ten characteristic relaxation times or longer. For the pure-polymer simulations, this characteristic time was taken as the polymer bond-orientational relaxation time, $\tau_{Rp}$, whereas for the probe-containing systems, it was taken as the probe rotational relaxation time, $\tau_{Rd}$.

We evaluate the translational and rotational dynamics of both the polymer and the probe molecules. To keep the notation concise, we define the dynamical correlation functions generally and append subscripts ($p$ for polymer monomers and $d$ for probe dimers) strictly to the extracted characteristic quantities, such as structural relaxation times ($\tau_{Sp}(k)$, $\tau_{Sd}(k)$), rotational relaxation times ($\tau_{Rp}$, $\tau_{Rd}$), and translational diffusion coefficients ($D_{Tp}$, $D_{Td}$).

The rotational dynamics is characterized by the second-order bond orientational time correlation function, $C(t)$:
\begin{equation}
C(t) = \left\langle P_2[\hat{\mathbf{u}}(0) \cdot \hat{\mathbf{u}}(t)] \right\rangle,
\end{equation}
where $\hat{\mathbf{u}}(t)$ denotes the unit bond vector at time $t$, $P_2(x) = (3x^2 - 1)/2$ is the second Legendre polynomial, and $\langle \cdots \rangle$ denotes an ensemble and time average. The rotational relaxation times extracted from this function are denoted as $\tau_{Rp}$ for the polymer and $\tau_{Rd}$ for the probe dimers. We also characterize finite-time rotational displacements using the cumulative angular displacement distribution,
\begin{equation}
G_s(\theta,t)=
\left\langle
\frac{1}{N_d}\sum_{j=1}^{N_d}
\delta\!\left[\theta-\Theta_j(t_0,t)\right]
\right\rangle ,
\end{equation}
where
\begin{equation}
\Theta_j(t_0,t)=
\sum_{n=0}^{M-1}
\cos^{-1}\!\left[
\hat{\mathbf{u}}_j(t_0+n\Delta t)\cdot
\hat{\mathbf{u}}_j(t_0+(n+1)\Delta t)
\right],
\end{equation}
with $t=M\Delta t$. Here, $\hat{\mathbf{u}}_j$ is the unit bond vector of dimer $j$, $\Delta t$ is the sampling interval, and the brackets denote averaging over dimers and time origins.

To quantify the translational dynamics, we compute the mean-squared displacement and the self-part of the intermediate scattering function ($F_s(k,t)$). The mean-squared displacement is given by:
\begin{equation}
\langle \Delta r^2(t) \rangle = \left\langle \frac{1}{N_{\alpha}} \sum_{j=1}^{N_{\alpha}} (\mathbf{r}_{j}(t) - \mathbf{r}_{j}(0))^2 \right\rangle,
\end{equation}
where $\mathbf{r}_{j}(t)$ denotes the position vector of particle $j$ at time $t$, and $N_{\alpha}$ is the number of polymer monomers or probe-dimer centers of mass included for species $\alpha$. The translational diffusion coefficient, $D_T$ (denoted as $D_{Tp}$ for the polymer and $D_{Td}$ for the dimer), is estimated by fitting the reliable long-time linear regime of the mean-squared displacement to the Einstein relation, $\langle \Delta r^2(t) \rangle = 6D_Tt$.

The spatial distribution of particle displacements and its corresponding structural relaxation are characterized by the self-part of the van Hove correlation function, $G_s(r,t)$, and its spatial Fourier transform, the self-intermediate scattering function, $F_s(k,t)$:
\begin{equation}
G_s(r, t) = \left\langle \frac{1}{N_{\alpha}} \sum_{j=1}^{N_{\alpha}} \delta \left(r - |\mathbf{r}_{j}(t) - \mathbf{r}_{j}(0)|\right) \right\rangle,
\end{equation}
\begin{equation}
F_s(k,t) = \left\langle \frac{1}{N_{\alpha}} \sum_{j=1}^{N_{\alpha}} \exp\left\{i\mathbf{k} \cdot \left[\mathbf{r}_{j}(t) - \mathbf{r}_{j}(0)\right]\right\} \right\rangle.
\end{equation}
We consider the wavevector magnitude $k = |\mathbf{k}|$ in the range from $\pi/\sigma$ to $4\pi/\sigma$. The structural relaxation times extracted from $F_s(k,t)$ are denoted as $\tau_{Sp}(k)$ and $\tau_{Sd}(k)$. For the probe dimers, both the mean-squared displacement and $F_s(k,t)$ are calculated based on the trajectories of their center of mass.

To characterize dynamic heterogeneity and its scale dependence, we employ three dynamic quantities.

First, the deviation of the single-particle mobility from normal Gaussian diffusion is evaluated by the non-Gaussian parameter, $\alpha_2(t)$:
\begin{equation}
\alpha_2(t) = \frac{3\left\langle \Delta r^4(t) \right\rangle}{5\left\langle \Delta r^2(t) \right\rangle^2} - 1.
\end{equation}

Second, the extent of spatially correlated structural relaxation in the bulk polymer is estimated using the overlap-based four-point dynamic susceptibility, $\chi_4(t)$:
\begin{equation}
\chi_4(t) = N_p \left[ \langle Q(t)^2 \rangle - \langle Q(t) \rangle^2 \right],
\end{equation}
where $N_p$ is the total number of polymer monomers, and the self overlap function is $Q(t) = \frac{1}{N_p} \sum_{j=1}^{N_p} w(|\mathbf{r}_{j}(t) - \mathbf{r}_{j}(0)|)$ with the step function $w(r) = 1$ if $r \le 0.3\sigma$, and $0$ otherwise. 

Finally, to characterize scale-dependent dynamic heterogeneity, we define the $k$-dependent dynamic susceptibility ($\chi_4(k,t)$) as the variance of the polymer's intermediate scattering function:
\begin{equation}
\chi_4(k,t) = N_p \left[ \langle |F_s(k,t)|^2 \rangle - \langle F_s(k,t) \rangle^2 \right].
\end{equation}

Error bars represent the standard error of the mean estimated from independent simulation trajectories. For fitted quantities, the error bars include fitting uncertainties and propagated errors.

\section{Results and Discussion}

\subsection{Interaction Dependence of Probe Rotational Relaxation Dynamics}

\begin{figure}
    \centering
    \includegraphics[width=\linewidth]{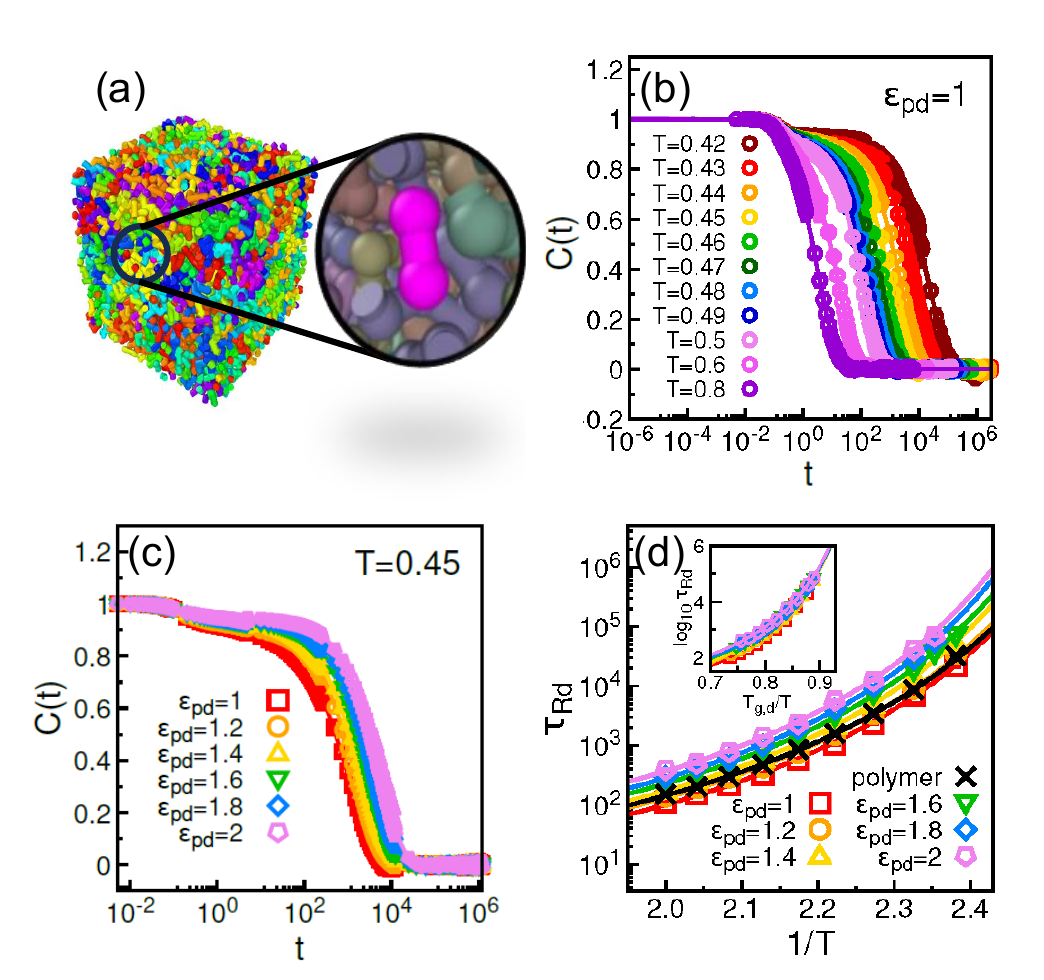}
    \caption{Interaction-dependent rotational relaxation dynamics of probe dimers. (a) A representative simulation snapshot of probe dimers (pink) dispersed within a polymer matrix (multi-colored beads). (b) The bond orientational time correlation function of probe dimers, \(C(t)\), for \(\varepsilon_{pd}=1\varepsilon\) across a range of temperatures (\(0.42 \leq T \leq 0.8\)). (c) \(C(t)\) at \(T=0.45\) for varying probe--polymer interaction strengths (\(1\varepsilon \leq \varepsilon_{pd} \leq 2\varepsilon\)). In panels (b) and (c), markers represent simulation results, and solid lines denote fits using a combination of exponential and stretched-exponential functions. (d) The rotational relaxation times of the probe dimers (colored symbols) and the bond-orientational relaxation time of the bulk polymer (black symbols) as a function of inverse temperature, \(1/T\). Solid lines denote VFT fits. For the probe data, the probe-specific prefactor \(\tau_{0,d}^{\mathrm{shift}}\), obtained from the vertical-shift analysis of the polymer VFT, is held fixed while \(B\) and \(T_0\) are fitted. The inset shows the corresponding Angell representation, in which the probe relaxation times are plotted against \(T_{g,d}/T\). The accessible simulation range corresponds to $T_{g,d}/T \le 0.9$.}
    \label{fig1}
\end{figure} 

To investigate how probe molecules report the glassy dynamics of a polymer matrix, we analyze the rotational relaxation of the probe dimers across various interaction strengths ($\varepsilon_{pd}$). The system-averaged dynamics of the host polymer matrix remains essentially unchanged by the presence of the dilute probe dimers over the interaction range examined (Figure S1 in the Supporting Information). For comparison, the bond orientational relaxation time of the bulk polymer ($\tau_{Rp}$) was independently obtained from a pure polymer matrix under identical conditions, and its corresponding correlation function $C(t)$ is shown in Figure S2 in the Supporting Information.

As shown in Figures~\ref{fig1}(b) and~\ref{fig1}(c), the bond orientational time correlation function of the probe dimers ($C(t)$) exhibits a typical two-step decay, shifting toward longer time scales upon cooling or with increasing $\varepsilon_{pd}$. We extract the rotational relaxation time of the dimers ($\tau_{Rd}$) by fitting the two-step decay to a combination of an exponential and a Kohlrausch-Williams-Watts (KWW) stretched exponential function, $C(t) = (1-A)\exp(-t/\tau_f) + A\exp[-(t/\tau_{Rd})^\beta]$\cite{Layding2026}. Here, $A$ represents the plateau value corresponding to the fraction of slowly relaxing correlations (the Debye-Waller factor analog), $\tau_f$ is the relaxation time for the fast local motion within the cage, and $\beta$ is the stretching exponent. The obtained stretching exponents ($\beta$) remain within a relatively narrow range over the supercooled regime for both the polymer and the dimer (Figure S3 in the Supporting Information).

Figure~\ref{fig1}(d) shows that $\tau_{Rd}$ increases with stronger probe--host interactions. For instance, at $T = 0.43$, $\tau_{Rd}$ for $\varepsilon_{pd}=2\varepsilon$ is approximately 1.7 times longer than that for $\varepsilon_{pd}=1\varepsilon$. A direct comparison of the absolute relaxation times reveals that no single $\varepsilon_{pd}$ enables the probe to reproduce the host dynamics across the entire temperature range. At $T = 0.5$, the polymer's relaxation time ($\tau_{Rp} \approx 150\tau$) corresponds to a probe interaction strength between $\varepsilon_{pd}=1.2\varepsilon$ ($\approx 140\tau$) and $1.4\varepsilon$ ($\approx 170\tau$). However, as the temperature decreases to $T = 0.43$, $\tau_{Rp}$ ($\approx 8600\tau$) better aligns with an interaction strength between $\varepsilon_{pd}=1\varepsilon$ ($\approx 7100\tau$) and $1.2\varepsilon$ ($\approx 9100\tau$). This implies that the probe-reported relaxation dynamics is sensitive to both the temperature and the probe--host interaction strength.

At high temperatures, the rotational relaxation follows approximately Arrhenius behavior, whereas a clear departure from the high-temperature Arrhenius trend develops upon cooling toward \(T \approx 0.5\) (Figure S4 in the Supporting Information). We therefore describe the low-temperature relaxation (\(T \leq 0.5\)) using the Vogel--Fulcher--Tammann (VFT) relation, \(\tau = \tau_0^{\mathrm{VFT}}\exp[B/(T-T_0)]\)\cite{Vogel1921,Fulcher1925,Tammann1926}. To assess how the VFT-based interpretation depends on the fitting protocol and to facilitate comparison with experimental rotational-probe analyses, we considered several parameter constraints. Experimental studies commonly examine whether probe relaxation preserves the temperature dependence of the host dynamics primarily through a vertical shift and subsequently use VFT fits to compare probe- and host-derived fragilities\cite{Kim2022}. For the fit shown in Figure~\ref{fig1}(d), a probe-specific prefactor, \(\tau_{0,d}^{\mathrm{shift}}\), is first obtained by fixing \(B_p\) and \(T_{0,p}\) to the polymer values and fitting only \(\tau_0\). The resulting \(\tau_{0,d}^{\mathrm{shift}}\) is then held fixed while \(B_d\) and \(T_{0,d}\) are fitted to the probe relaxation data. The VFT parameters and their standard errors for all fitting protocols are summarized in Table S1 in the Supporting Information.

To examine the VFT description more systematically, we compared several constrained fitting protocols. Figure S5 in the Supporting Information shows the alternative constrained fits, while the probe-specific \(\tau_{0,d}^{\mathrm{shift}}\)-fixed fit is shown in Figure~\ref{fig1}(d). In the vertical-shift fit, \(B_p\) and \(T_{0,p}\) are fixed to the polymer values and only \(\tau_0\) is varied. We also considered a one-parameter fit in which \(\tau_{0,p}\) and \(B_p\) are fixed and only \(T_0\) is varied, a common-prefactor fit in which \(\tau_{0,p}\) is fixed while \(B\) and \(T_0\) are fitted, and the two-step fit in which the probe-specific \(\tau_{0,d}^{\mathrm{shift}}\) obtained from the vertical-shift analysis is fixed while \(B\) and \(T_0\) are fitted. The common-prefactor and two-step fits provide particularly good descriptions of the simulation data, with adjusted \(R^2\) values above 0.997 over the entire range of \(\varepsilon_{pd}\). The two one-parameter descriptions also provide good fits, although their adjusted \(R^2\) values show a somewhat stronger dependence on \(\varepsilon_{pd}\) and remain above 0.96 throughout the interaction range (Figure S6 in the Supporting Information). For comparison, experimental rotational-probe measurements analyzed using analogous one-parameter tests showed adjusted \(R^2\) values close to unity, with an average of \(0.996\pm0.006\) computed from the data reported in Kim et al.\cite{Kim2022}, supporting an approximately vertical-shift-like relation between probe and host relaxation. Thus, the high goodness of fit alone does not uniquely distinguish among the VFT parameterizations in our simulations. 

We next examined how the apparent glass-transition temperature and fragility depend on the VFT fitting protocol. We first evaluated the probe fragility at the probe-specific operational glass-transition temperature, $T_{g,d}$, defined by $\tau_{Rd}(T_{g,d})=10^{14}\tau$. The corresponding apparent fragility was evaluated from the VFT slope as \(m_f=B T_g/[\ln(10)(T_g-T_0)^2]\). For the polymer reference, the same criterion gives $T_{g,p}=0.3660 \pm 0.0022$ and $m_{f,p}=200.1 \pm 11.7$. When $B_p$ and $T_{0,p}$ are fixed and only $\tau_0$ is fitted, the apparent fragility decreases moderately with increasing $\varepsilon_{pd}$, whereas fixing $\tau_{0,p}$ and fitting $B$ and $T_0$ produces a stronger decrease (Figures S7(a) and S7(b) in the Supporting Information). In contrast, when the probe-specific $\tau_{0,d}^{\mathrm{shift}}$ obtained from the vertical-shift analysis is fixed and $B_d$ and $T_{0,d}$ are subsequently fitted, the interaction dependence is much weaker (Figure S7(c) in the Supporting Information). The corresponding Angell representation in the inset of Figure~\ref{fig1}(d) shows modest differences in the normalized temperature dependence among the probe systems. Under this latter constraint, the probe-derived fragilities differ from the polymer reference value by approximately 3.8--10.4\% over the interaction range examined. Experimental rotational-probe analyses using a related VFT procedure reported deviations within approximately 10\% of the host fragility\cite{Kim2022}. Although the experimental fragilities were evaluated using an independently determined host $T_g$, rather than a probe-derived $T_{g,d}$, the magnitude of the probe--host difference obtained in our simulations is therefore of a comparable scale.

The apparent fragility, however, is sensitive to both the VFT parameter constraints and the temperature at which the VFT slope is evaluated. This is particularly evident when the same fitted VFT parameters are evaluated at the common polymer glass-transition temperature, $T_{g,p}$, rather than at $T_{g,d}$ (Figure S8 in the Supporting Information). Although $T_{g,d}$ changes only modestly with $\varepsilon_{pd}$, the fragility contains the factor $(T_g-T_0)^{-2}$, so relatively small differences in $T_g$ and fitted $T_0$ can be strongly amplified. As a result, evaluating the slope at $T_{g,p}$ produces substantially larger apparent fragilities and uncertainties at stronger interactions. The VFT parameters are also strongly correlated, as summarized in Table S1, further limiting the uniqueness of the inferred fragility. Importantly, varying the upper fitting temperature from $T_{\max}=0.50$ to $0.47$ does not qualitatively alter the trends obtained within the fixed-$\tau_{0,p}$ or fixed-$B_p$,$T_{0,p}$ protocols (Figures S9 and S10 in the Supporting Information), indicating that these protocol-dependent trends are not an artifact of a particular fitting window. Their quantitative magnitude, however, remains sensitive to the choice of VFT constraint and \(T_g\) definition.

Taken together, these analyses show that the VFT-derived probe fragility is not uniquely constrained by the relaxation data over the accessible temperature range in this study. Although several VFT parameterizations describe the simulated data comparably well, the resulting fragilities depend appreciably on the fitting protocol and on the definition of \(T_g\). In our simulation data, a particular VFT analysis can yield a probe-derived fragility close to the polymer reference value, with deviations comparable in magnitude to those reported experimentally. Accordingly, we do not assign physical significance to the precise magnitude or interaction dependence of the VFT-derived fragility, and do not use it as a primary measure of probe--host coupling in the present work.

\subsection{Interaction-Dependent Probe--Host Relaxation Scaling}

\begin{figure}
    \centering
    \includegraphics[width=\linewidth]{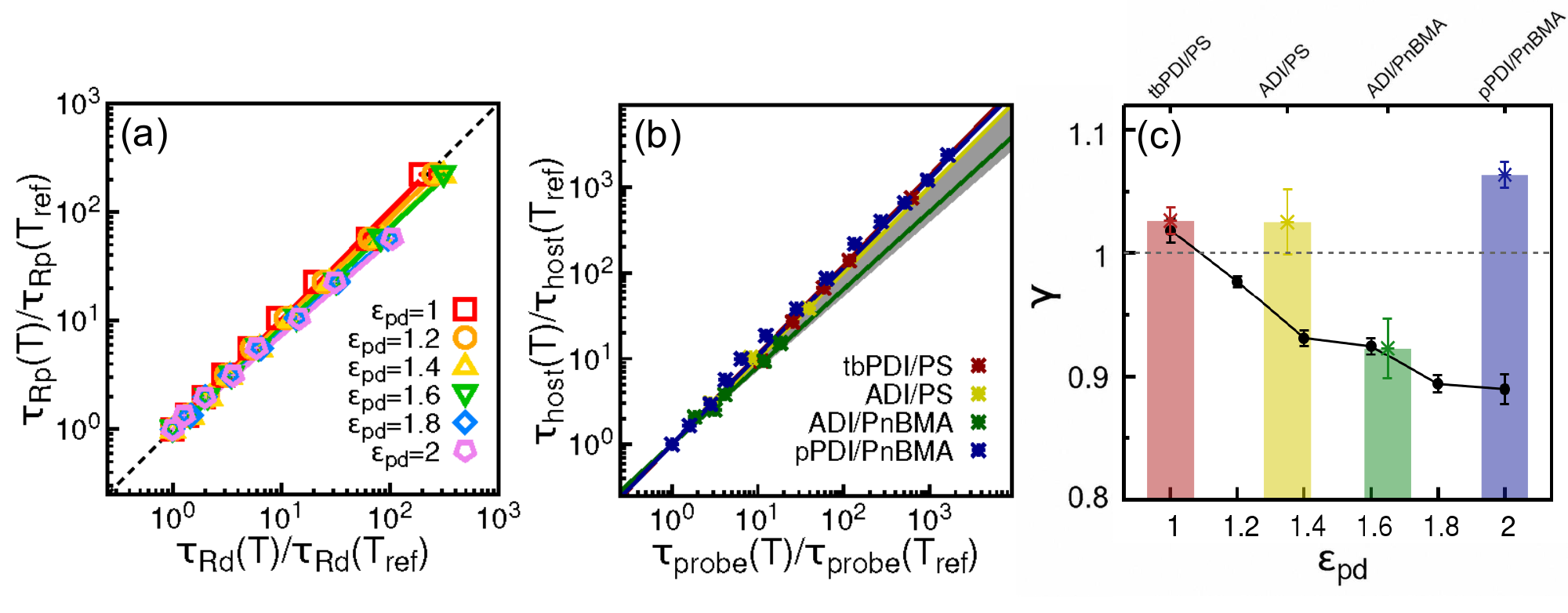}
\caption{Normalized probe--host relaxation scaling in simulations and experiments. (a) Relative polymer bond-orientational relaxation time, $\tau_{Rp}(T)/\tau_{Rp}(T_{\mathrm{ref}})$, plotted against the corresponding relative probe rotational relaxation time, $\tau_{Rd}(T)/\tau_{Rd}(T_{\mathrm{ref}})$, for $\varepsilon_{pd}=1\varepsilon$--$2\varepsilon$. Here, $T_{\mathrm{ref}}=0.5$. Solid lines denote fits to the fractional scaling relation $\tau_{Rp}(T)/\tau_{Rp}(T_{\mathrm{ref}})=[\tau_{Rd}(T)/\tau_{Rd}(T_{\mathrm{ref}})]^{\gamma}$, and the diagonal dashed line denotes one-to-one scaling ($\gamma=1$). (b) Corresponding normalized host and probe relaxation times for the experimental probe--polymer systems reported by Kim et al.~\cite{Kim2022}: tbPDI/PS, ADI/PS, ADI/PnBMA, and pPDI/PnBMA. For each experimental system, $T_{\mathrm{ref}}$ is taken as the highest temperature for which probe rotational relaxation data are reported. For the experimental data, the corresponding high-temperature reference points are 333.2 K for pPDI/PnBMA, 305.5 K for ADI/PnBMA, 390 K for tbPDI/PS, and 377.9 K for ADI/PS. The corresponding host relaxation times are evaluated from the VFT representation of the dielectric relaxation data reported in the same work, and both the probe and host relaxation times are normalized by their values at $T_{\mathrm{ref}}$. The gray shaded region indicates the range of scaling behavior spanned by the simulations in panel (a). (c) Fractional scaling exponent $\gamma$ as a function of $\varepsilon_{pd}$ for the simulations (black circles), together with the corresponding values obtained from the experimental systems (colored crosses). The experimental values are identified by the auxiliary top axis and are shown for comparison only. The horizontal dashed line at $\gamma=1$ denotes identical relative temperature dependence of the probe and host relaxation. The scaling exponents were obtained by linear regression in log–log space with the intercept fixed at zero. Error bars denote the standard errors of the fitted slopes.}
\label{fig2}
\end{figure}

To characterize the probe--host correspondence directly within the simulated temperature range, we compare their normalized relaxation times. To separate differences in temperature dependence from differences in absolute relaxation time, both the probe and polymer relaxation times are normalized by their values at a reference temperature, \(T_{\mathrm{ref}}=0.5\), and related through

$$ \frac{\tau_{Rp}(T)}{\tau_{Rp}(T_{\mathrm{ref}})} = \left[ \frac{\tau_{Rd}(T)}{\tau_{Rd}(T_{\mathrm{ref}})} \right]^\gamma . $$

Here, \(\gamma=1\) corresponds to identical relative temperature dependence of the probe and polymer relaxation, equivalent to an exact temperature-independent vertical shift between their unnormalized relaxation times. Values of \(\gamma\neq1\) quantify deviations from this limiting behavior without requiring extrapolation to \(T_g\).

As shown in Figure~\ref{fig2}(a), the weakly interacting probe (\(\varepsilon_{pd}=1\varepsilon\)) closely follows one-to-one scaling with the polymer relaxation, with \(\gamma\approx1.02\). As \(\varepsilon_{pd}\) increases, \(\gamma\) decreases systematically, reaching approximately 0.89 at \(\varepsilon_{pd}=2\varepsilon\) [Figure~\ref{fig2}(c)].  With the scaling relation defined above, \(\gamma<1\) indicates that the probe relaxation exhibits a stronger relative temperature dependence than the polymer relaxation. Because the probe geometry and host model are unchanged across these simulations, this systematic variation reflects the effect of changing the probe--host affinity within the present model. We also repeated this analysis using relaxation times defined operationally as the times at which the corresponding correlation functions decay to 0.2. The same qualitative interaction dependence of $\gamma$ is obtained (Figure S11 in the Supporting Information), indicating that the observed scaling trend is not specific to the procedure used to determine $\tau_{Rd}$ and $\tau_{Rp}$.

For experimental context, we applied the same normalized scaling analysis to the reported rotational-probe relaxation times and the corresponding host dielectric-relaxation times evaluated from the reported VFT parameters for selected PS and PnBMA systems\cite{Kim2022} [Figure~\ref{fig2}(b)]. These systems provide temperature-dependent probe and host relaxation data over a sufficient range and allow different probes to be compared within the same host polymer matrix. Most of the experimental systems remain close to the one-to-one limit: tbPDI/PS and ADI/PS exhibit \(\gamma\) values near unity, whereas ADI/PnBMA shows a sub-unity value of approximately 0.92 and pPDI/PnBMA exhibits a modest deviation in the opposite direction [Figure~\ref{fig2}(c)]. Thus, the experimental systems examined here predominantly exhibit an approximately host-like temperature dependence, while also showing that small probe-specific departures from exact one-to-one scaling can occur.

The experimental results provide a useful reference for assessing the magnitude of the fractional deviations observed in the simulations. The simulation results overlap with the experimental range over part of the interaction window, while the highest values of \(\varepsilon_{pd}\) produce somewhat larger sub-unity deviations. We do not attempt a direct mapping between individual experimental systems and specific values of \(\varepsilon_{pd}\), because probe size, molecular shape, chemical functionality, specific interactions, and host chemistry vary simultaneously in experiments. In the simulations, these factors are held fixed while \(\varepsilon_{pd}\) is varied, allowing the effect of probe--host affinity on the scaling behavior to be examined under controlled conditions.

Probe size provides an additional source of variation in the probe--host dynamical correspondence. To examine whether the interaction-dependent scaling is specific to the original dimer size, we performed additional simulations with larger dimer probes characterized by \(s=1.5\) and 2, where \(s\) denotes the LJ diameter of each probe bead relative to the polymer-monomer diameter (\(\sigma_{dd}=s\sigma\)). The probe--polymer diameter was adjusted consistently using the same mixing rule as in the original model. For each size, we considered \(\varepsilon_{pd}=1\varepsilon\) and \(2\varepsilon\). As shown in Figure S12 in the Supporting Information, the absolute value of \(\gamma\) depends appreciably on probe size: for \(s=1.5\), \(\gamma\) decreases from approximately 0.94 to 0.78 as \(\varepsilon_{pd}\) increases from $1\varepsilon$ to $2\varepsilon$, while for \(s=2\) it decreases from approximately 0.93 to 0.77. The larger probes therefore exhibit stronger fractional deviations even at weak interaction, but the decrease in \(\gamma\) with increasing probe--host attraction is retained for all three probe sizes examined. These additional calculations also illustrate the limits of the present minimal probe model. The two-bead dimer provides only a minimal representation of experimental fluorescent probes, for which molecular size and architecture can both influence the reported dynamics\cite{Zhu2023}. The values of \(s\) should not be mapped directly onto particular molecular probes. Rather, the size variation provides a controlled test showing that both probe size and probe--host affinity influence the quantitative scaling, while the interaction-dependent change in \(\gamma\) is not restricted to a single probe dimension.

\subsection{Wavevector-Dependent Correspondence between Probe and Host Relaxation}

\begin{figure}
    \centering
    \includegraphics[width=\linewidth]{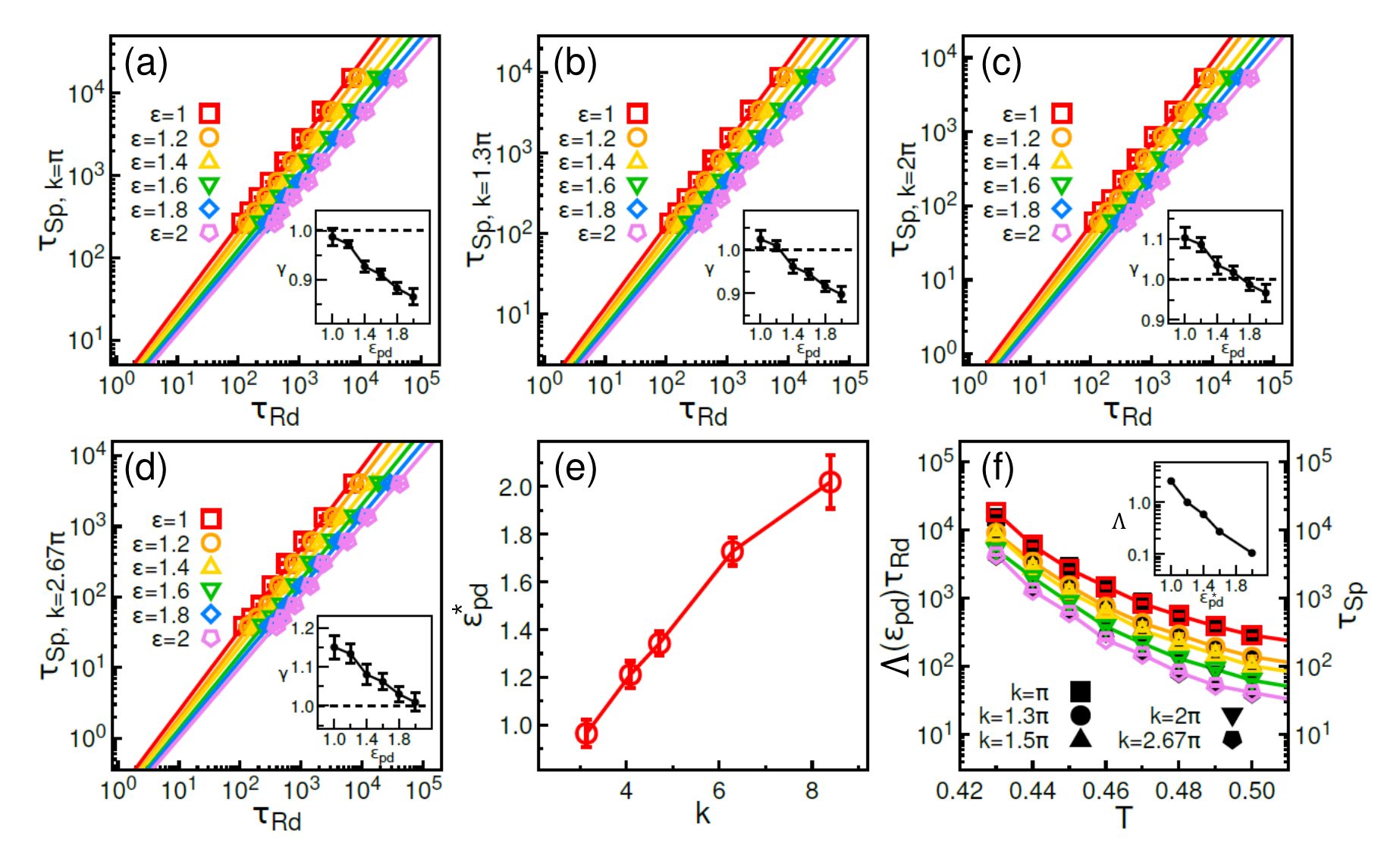}
\caption{Wavevector-dependent correspondence between probe rotational and polymer structural relaxation. 
(a--d) Correlations between the rotational relaxation time of the dimer, $\tau_{Rd}$, and the structural relaxation time of the bulk polymer, $\tau_{Sp}(k)$, at representative wavevectors: (a) $k=\pi/\sigma$, (b) $k=1.3\pi/\sigma$, (c) $k=2\pi/\sigma$, and (d) $k=2.67\pi/\sigma$. Markers represent simulation results, and solid lines denote fits to $\tau_{Sp}(k)=\Lambda\tau_{Rd}^{\gamma}$. Insets show the resulting scaling exponent $\gamma$ as a function of $\varepsilon_{pd}$. Black symbols denote the fitted $\gamma$ values, solid lines denote linear fits to the $\gamma$--$\varepsilon_{pd}$ relation, and horizontal dashed lines indicate $\gamma=1$. 
(e) Effective matching interaction strength, $\varepsilon_{pd}^{*}(k)$, as a function of wavevector $k$. For each $k$, $\varepsilon_{pd}^{*}(k)$ is obtained from a linear fit of the $\gamma$--$\varepsilon_{pd}$ relation as the interaction strength at which $\gamma=1$; representative fits are shown in the insets of panels (a--d). The solid line denotes a linear fit to $\varepsilon_{pd}^{*}(k)$. (f) Comparison of the scaled probe rotational relaxation times, $\Lambda(\varepsilon_{pd})\tau_{Rd}$, with the polymer structural relaxation times, $\tau_{Sp}(k)$, at representative wavevectors. For each wavevector, the probe data are shown for the discrete simulated interaction strength $\varepsilon_{pd}$ closest to the effective matching value $\varepsilon_{pd}^{*}(k)$ obtained in (e). Colored open markers represent the scaled probe relaxation times, and black markers represent $\tau_{Sp}(k)$. The marker colors for each $\varepsilon_{pd}$ follow those used in panels (a--d). The inset shows the corresponding scaling prefactor $\Lambda$ used to rescale the probe relaxation times as a function of $\varepsilon_{pd}$. The matching condition $\gamma=1$ denotes identical temperature dependence, apart from the temperature-independent prefactor $\Lambda$, and is used here as an operational measure of the wavevector-dependent probe--host correspondence.}
\label{fig3}
\end{figure}

To examine how the temperature dependence of probe rotational relaxation compares with the wavevector-dependent structural relaxation of the host, we compare \(\tau_{Rd}\) with the polymer structural relaxation time, \(\tau_{Sp}(k)\), at different wavevectors. The polymer structural relaxation time ($\tau_{Sp}(k)$) was extracted from the polymer self-intermediate scattering function $F_s(k,t)$ using the same relaxation criterion employed for the rotational correlation function $C(t)$ (Figure S13 in the Supporting Information).

To compare the intrinsic temperature dependence of the two dynamic processes, we employ a fractional dynamic mapping relation: $\tau_{Sp}(k, T) = \Lambda \cdot [\tau_{Rd}(\varepsilon_{pd}, T)]^\gamma$. In this relation, the prefactor $\Lambda$ accounts for the difference in the absolute time scales, while the exponent \(\gamma\) characterizes their relative temperature scaling. Specifically, \(\gamma=1\) indicates that the probe rotational relaxation and the host structural relaxation at a given \(k\) exhibit the same temperature dependence, apart from the temperature-independent prefactor \(\Lambda\), over the simulated temperature range.

Figure~\ref{fig3}(a)--(d) shows the relationship between \(\tau_{Sp}(k)\) and \(\tau_{Rd}\) for various interaction strengths at representative wavevectors. The insets show the extracted scaling exponent \(\gamma\) as a function of \(\varepsilon_{pd}\). By linearly fitting the \(\gamma\)--\(\varepsilon_{pd}\) relation, we identify an effective matching interaction strength, \(\varepsilon_{pd}^{*}(k)\), defined operationally as the interaction strength at which \(\gamma=1\) for each wavevector \(k\). Figure~\ref{fig3}(e) shows that \(\varepsilon_{pd}^{*}(k)\) increases approximately linearly with \(k\). Equivalently, increasing \(\varepsilon_{pd}\) shifts the host wavevector whose structural relaxation exhibits the closest temperature dependence to the probe relaxation toward larger \(k\). We interpret this relation as a phenomenological wavevector-dependent correspondence.


Figure~\ref{fig3}(f) further illustrates the wavevector-dependent correspondence by comparing the scaled probe rotational relaxation time with the polymer structural relaxation time at representative wavevectors. For each \(k\), the probe data are taken from the discrete simulated interaction strength closest to the corresponding \(\varepsilon_{pd}^{*}(k)\). The scaled probe and polymer relaxation times then exhibit similar temperature dependences over the simulated range. As \(\varepsilon_{pd}\) increases, the corresponding effective matching wavevector shifts from approximately \(k=\pi/\sigma\) at weak interaction toward larger wavevectors, reaching \(k=2\pi/\sigma\) and \(2.67\pi/\sigma\) for the higher interaction strengths illustrated here. Although \(2\pi/k\) provides a nominal real-space wavelength associated with each wavevector, this correspondence does not imply that the probe rotation is uniquely coupled to or samples a specific microscopic length \(2\pi/k\). The matching wavevector provides an operational description of how the temperature dependence of probe relaxation compares with the wavevector-resolved host dynamics.

\begin{figure}
    \centering
    \includegraphics[width=0.8\linewidth]{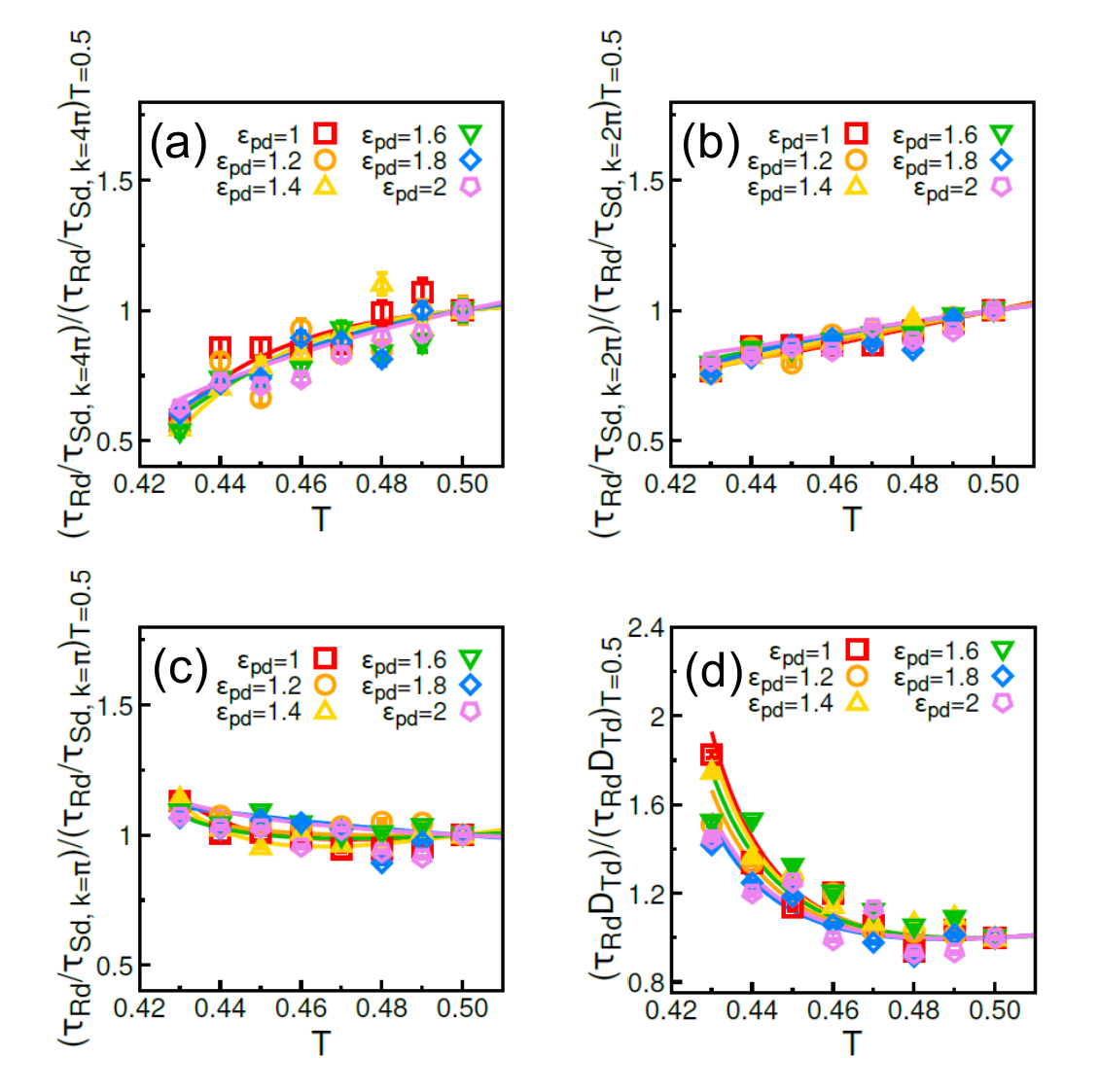}
    \caption{Decoupling between rotational and translational dynamics of the probe dimers. (a--c) Temperature dependence of the normalized ratio between the rotational relaxation time ($\tau_{Rd}$) and the structural relaxation time ($\tau_{Sd}$) of the probe dimers at different length scales: (a) $k = 4\pi/\sigma$, (b) $k = 2\pi/\sigma$, and (c) $k = \pi/\sigma$. (d) Temperature dependence of the normalized quantity $\tau_{Rd}D_{Td}$. All quantities in (a--d) are normalized by their respective values at $T = 0.5$. Markers represent simulation results, and solid lines represent ratios obtained from the corresponding VFT fits.
}
    \label{fig4}
\end{figure}

To further characterize the relation between probe rotational and translational relaxation across different wavevectors, we examine their decoupling behavior. Figure~\ref{fig4}(a)--(c) shows the temperature dependence of the normalized ratio $\tau_{Rd}/\tau_{Sd}$ at different wavevectors. The probe structural relaxation time $\tau_{Sd}(k)$ was similarly obtained from the probe self-intermediate scattering function $F_s(k,t)$ using the same relaxation criterion (Figures S14--S16 in the Supporting Information). The decoupling behavior depends strongly on the length scale used to define the probe structural relaxation. For wavevectors larger than $k=\pi/\sigma$, the ratio decreases upon cooling, whereas at $k=\pi/\sigma$ the ratio remains nearly temperature independent over the entire supercooled regime.

The different temperature trends observed at different wavevectors show that the relation between probe rotational and translational relaxation depends on the length scale over which translational relaxation is measured. At relatively large wavevectors ($k=4\pi/\sigma$), the translational relaxation probes highly localized motions associated with intra-cage fluctuations. In this regime, the localized translational relaxation associated with intra-cage confinement slows down more strongly than the rotational relaxation upon cooling. In contrast, the near temperature-independent ratio at \(k=\pi/\sigma\) indicates that rotational relaxation and translational relaxation measured at this wavevector exhibit closely similar temperature dependences. The corresponding nominal wavelength, \(2\pi/k=2\sigma\), is comparable to the dimer body length, providing a useful geometric reference for this correspondence. Overall, the results demonstrate a pronounced wavevector dependence in the relative temperature evolution of rotational and translational relaxation, ranging from more local translational relaxation at large \(k\) to longer-length-scale transport at smaller \(k\).

Figure~\ref{fig4}(d) shows the temperature dependence of the normalized quantity $\tau_{Rd}D_{Td}$. The translational diffusion coefficient $D_{Td}$ was estimated from the long-time behavior of the probe mean-squared displacement (Figure S17 in the Supporting Information). Unlike the finite wavevector structural relaxation time $\tau_{Sd}(k)$, the diffusion coefficient $D_{Td}$ reflects the long time, long length scale transport of the probe dimers. The increase of $\tau_{Rd}D_{Td}$ upon cooling therefore indicates the progressive decoupling between rotational relaxation and macroscopic translational diffusion in the supercooled regime, which is commonly associated with hopping-assisted translational transport in glass-forming liquids\cite{Mazza2006,Becker2006,Liu2015}. Because the probe dimers considered here are relatively small, hopping-like translational motion may contribute to their long-range transport. The interaction dependence of \(\tau_{Rd}D_{Td}\) further indicates that the degree of rotational--translational decoupling varies with probe--host affinity.

\subsection{Interaction-Dependent Sampling of Heterogeneous Host Dynamics}

\begin{figure}
    \centering
    \includegraphics[width=\linewidth]{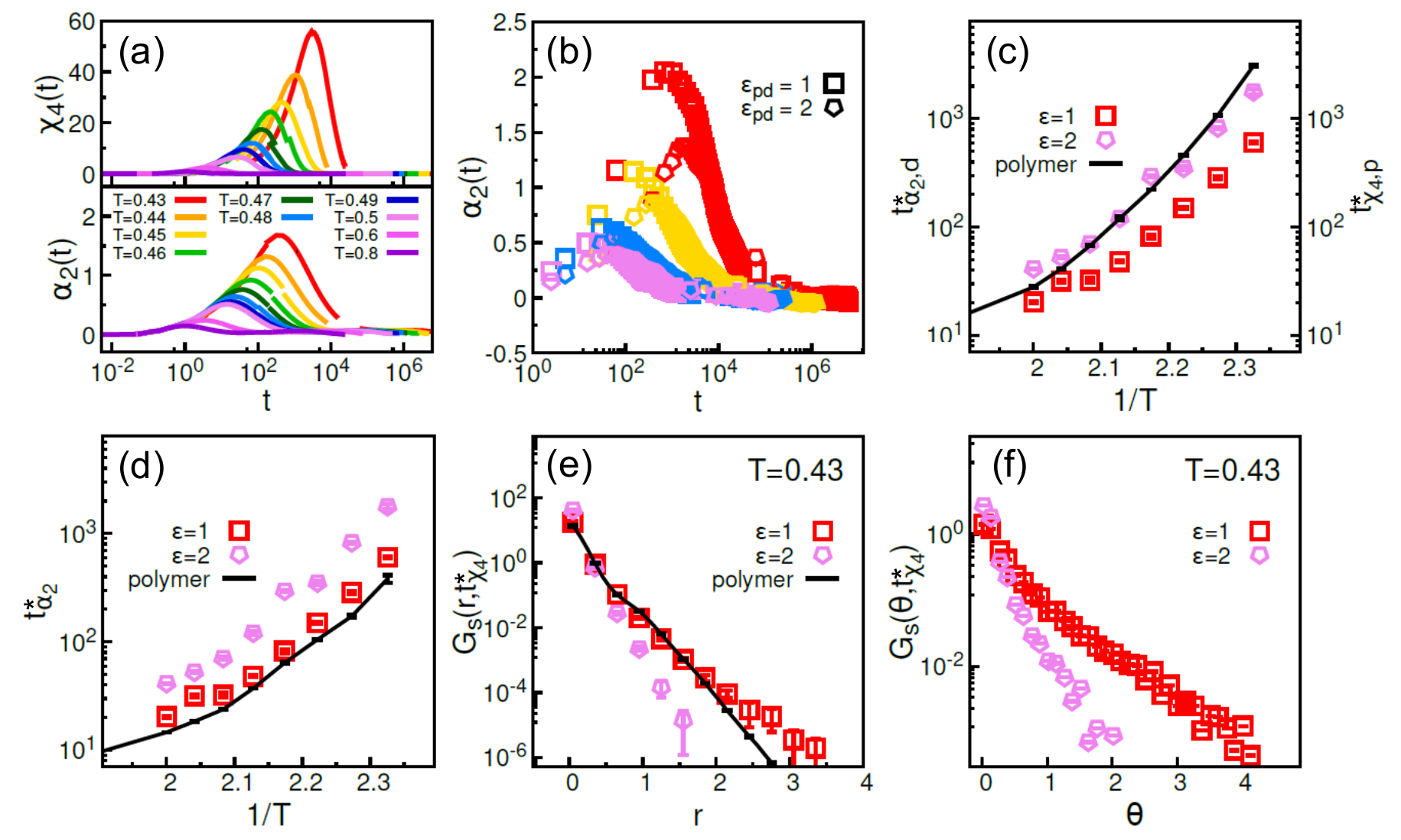}
    \caption{Dynamic heterogeneity and interaction-dependent transport of probe dimers. 
    (a) Time evolution of the host four-point dynamic susceptibility $\chi_4(t)$ (top panel) and the host non-Gaussian parameter $\alpha_2(t)$ (bottom panel). 
    (b) Time evolution of $\alpha_2(t)$ for the probe dimers at different interaction strengths ($\varepsilon_{pd}=1\varepsilon$ (squares) and $2\varepsilon$ (inverted pentagons)). Temperature colors correspond to those used in panel (a). Error bars are omitted for clarity; corresponding results including error bars are provided in Figure S18 of the Supporting Information. 
    (c) Comparison between the peak time of the host four-point dynamic susceptibility ($t_{\chi_4}^{*}$, black) and the peak times of the probe non-Gaussian parameter ($t_{\alpha_2}^{*}$) for $\varepsilon_{pd}=1\varepsilon$ (red squares) and $2\varepsilon$ (magenta inverted pentagons) as a function of inverse temperature ($1/T$). 
    (d) Peak times of the non-Gaussian parameter ($t_{\alpha_2}^{*}$) as a function of inverse temperature ($1/T$) for the bulk polymer (black), probes with $\varepsilon_{pd}=1\varepsilon$ (red squares), and probes with $\varepsilon_{pd}=2\varepsilon$ (magenta inverted pentagons). Characteristic peak times in (c) and (d) were determined by quadratic fitting near the local maxima. 
    (e,f) Translational and rotational displacement distributions evaluated at the characteristic dynamic-heterogeneity time scale. 
    (e) Self-part of the van Hove correlation function ($G_s(r,t)$) evaluated at the peak time of the host four-point dynamic susceptibility ($t=t_{\chi_4}^{*}$) at $T=0.43$. 
    (f) Angular displacement distribution function ($G_s(\theta,t)$) evaluated at the same time ($t=t_{\chi_4}^{*}$) for $\varepsilon_{pd}=1\varepsilon$ (red squares) and $2\varepsilon$ (magenta inverted pentagons).}
    \label{fig5}
\end{figure}

To characterize the microscopic dynamical environments sampled by the probe, we examine the dynamic heterogeneity of both the bulk polymer and the probe dimers. For the bulk polymer, two characteristic time scales can be identified from the four-point dynamic susceptibility $\chi_4(t)$ and the non-Gaussian parameter $\alpha_2(t)$. The peak time of $\alpha_2(t)$ ($t_{\alpha_2}^{*}$) characterizes the time scale of enhanced single-particle heterogeneous motion, whereas the peak time of $\chi_4(t)$ ($t_{\chi_4}^{*}$) corresponds to the time scale of collective structural rearrangement of the surrounding matrix\cite{xu2016,xu2022}. As shown in Figure~\ref{fig5}(a), the peak in $\alpha_2(t)$ precedes that of $\chi_4(t)$, indicating that heterogeneous single-particle motion emerges before the collective relaxation of the host matrix.

Figure~\ref{fig5}(b) shows the time evolution of the non-Gaussian parameter $\alpha_2(t)$ for the probe dimers at different interaction strengths. The probe with $\varepsilon_{pd}=1\varepsilon$ exhibits substantially larger non-Gaussian peaks than both the bulk polymer and the probe with $\varepsilon_{pd}=2\varepsilon$, while the latter remains much closer to the host response. The characteristic time scales also show a clear interaction dependence. As shown in Figure~\ref{fig5}(c), $t_{\alpha_2}^{*}$ for $\varepsilon_{pd}=2\varepsilon$ lies closer to the host collective-heterogeneity time scale $t_{\chi_4}^{*}$, whereas Figure~\ref{fig5}(d) shows that the characteristic time for $\varepsilon_{pd}=1\varepsilon$ is closer to the host single-particle heterogeneity time $t_{\alpha_2}^{*}$. These local transport signatures characterize a different aspect of the probe dynamics from the normalized rotational-relaxation scaling discussed above. In particular, the enhanced non-Gaussian motion for $\varepsilon_{pd}=1\varepsilon$ can coexist with the nearly one-to-one probe--host rotational relaxation scaling ($\gamma\approx1$) observed in Figure~\ref{fig2}.

The interaction dependence of these characteristic time scales suggests that changing the probe--host affinity modifies the local dynamical environments explored by the probe. At $\varepsilon_{pd}=1\varepsilon$, the probe exhibits stronger heterogeneous translational motion and can undergo relatively large displacements within the host matrix. At \(\varepsilon_{pd}=2\varepsilon\), these large displacements are suppressed and the probe remains more localized within its surrounding polymer environment over the same observation time scale. Thus, the interaction dependence is manifested not only in the average relaxation time but also in the distribution and characteristic time scales of probe motion.

To examine this behavior directly at the time of maximum collective heterogeneity, we analyze the translational and rotational displacement distributions at $t=t_{\chi_4}^{*}$. Figure~\ref{fig5}(e) shows that the self-part of the van Hove correlation function $G_s(r,t)$ for the $\varepsilon_{pd}=1\varepsilon$ probe closely follows that of the bulk polymer at short displacements. The two distributions begin to diverge at approximately $r\approx2\sigma$, comparable to the dimer body length, beyond which the probe exhibits an enhanced long-displacement tail. This behavior reflects an increased probability of large translational displacements within the heterogeneous matrix. In contrast, the $\varepsilon_{pd}=2\varepsilon$ probe remains considerably more localized than the bulk polymer at the same time scale.

Figure~\ref{fig5}(f) shows the corresponding angular displacement distributions $G_s(\theta,t)$. The $\varepsilon_{pd}=1\varepsilon$ probe exhibits broader angular displacements than the $\varepsilon_{pd}=2\varepsilon$ probe, indicating greater rotational mobility over the same observation window. The difference between the translational and rotational displacement distributions suggests that rotational motion is more strongly constrained by the local geometry of the surrounding polymer than large translational excursions. This distinction may contribute to the rotational--translational decoupling observed in Figure~\ref{fig4}.

\begin{figure}
    \centering
    \includegraphics[width=0.8\linewidth]{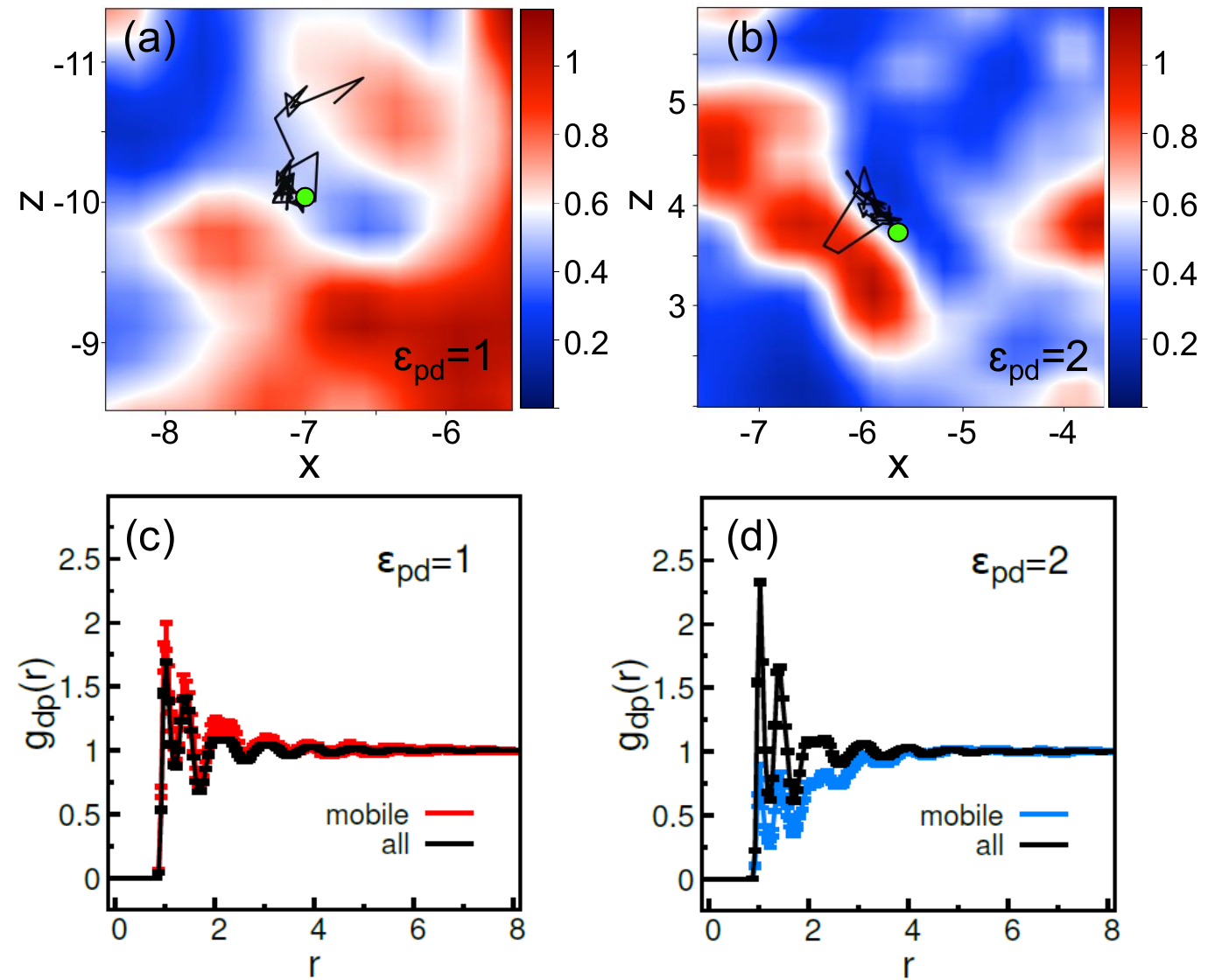}
    \caption{Interaction-dependent sampling of dynamic heterogeneity by probe dimers. 
    (a,b) Two-dimensional spatial mobility maps of the host polymer matrix with representative probe-dimer trajectories for (a) $\varepsilon_{pd}=1\varepsilon$ and (b) $\varepsilon_{pd}=2\varepsilon$ at $T=0.43$. The maps are projected onto the $x$--$z$ plane and show a spatial slice of thickness $\pm2\sigma$ centered on each probe dimer. The green circle marks the starting point of each probe trajectory. The colormap represents host monomer mobility, where the upper limit of the color scale corresponds to the mobility threshold defining the top 5\% most mobile monomers. Both the host mobility maps and the probe trajectories are evaluated over the same time window ($t=t_{\chi_4}^{*}$). 
    (c,d) Radial distribution functions $g_{dp}(r)$ between the probe dimers and the top 5\% most mobile host monomers, compared with the corresponding total host-monomer distributions, for (c) $\varepsilon_{pd}=1\varepsilon$ and (d) $\varepsilon_{pd}=2\varepsilon$.}
    \label{fig6}
\end{figure}

To examine how probe--host affinity affects the local environments explored by the probe, we compare representative trajectories for $\varepsilon_{pd}=1\varepsilon$ and $2\varepsilon$ with host mobility maps evaluated over the same characteristic time window, $t=t_{\chi_4}^{*}$, at $T=0.43$. As shown in Figures~\ref{fig6}(a) and~\ref{fig6}(b), the representative $\varepsilon_{pd}=1\varepsilon$ trajectory undergoes a relatively large translational excursion and passes through dynamically active regions of the host matrix. In contrast, the representative $\varepsilon_{pd}=2\varepsilon$ probe remains substantially more localized over the same observation window. These trajectories provide a qualitative illustration of the interaction-dependent local environments, which we characterize statistically below.

For a statistical characterization, we compare the radial distribution functions $g_{dp}(r)$ between the probe dimers and the top 5\% most mobile host monomers with the corresponding total host-monomer distributions (Figures~\ref{fig6}(c) and~\ref{fig6}(d)):
\[
g_{dp}^{\mathrm{mob}}(r)
=
\frac{1}{N_d \rho_{\mathrm{mob}} 4\pi r^2 \Delta r}
\left\langle N_{dp}^{\mathrm{mob}}(r,r+\Delta r) \right\rangle,
\]
where $N_d$ is the number of probe dimers, $\rho_{\mathrm{mob}}$ is the number density of the top 5\% most mobile host monomers, and $N_{dp}^{\mathrm{mob}}(r,r+\Delta r)$ is the number of mobile host monomers within a spherical shell between $r$ and $r+\Delta r$ from the probe-dimer center of mass. The total-monomer RDF, $g_{dp}^{\mathrm{all}}(r)$, was calculated using the same expression with all host monomers and the corresponding total monomer number density $\rho_{\mathrm{all}}$.

For \(\varepsilon_{pd}=1\varepsilon\), the mobile-monomer distribution nearly overlaps with the total-monomer distribution. Thus, the local environments sampled by this probe show little preferential enrichment or depletion of the most mobile host monomers. This relatively unbiased local sampling is compatible with the nearly one-to-one rotational relaxation scaling observed for this interaction strength (\(\gamma\approx1\); Figure~\ref{fig2}). In contrast, for \(\varepsilon_{pd}=2\varepsilon\), the mobile-monomer distribution lies substantially below the total-monomer distribution, indicating that the higher-affinity probe is statistically less exposed to the most dynamically active local environments. This change in the sampled local environment accompanies the sub-unity probe--host scaling observed at stronger interaction. The interaction-dependent differences in \(g_{dp}(r)\) therefore provide a real-space signature that probe--host affinity modifies the local dynamical environments sampled by the probe.

Previous studies of probe dynamics in glass-forming liquids have emphasized probe size as an important factor controlling how molecular probes sample heterogeneous host dynamics\cite{Liu2015,Li2023}. Small probes can exhibit hopping-like motion and enhanced translational heterogeneity, whereas larger probes tend to average over their surrounding environment. The present results show that, even at fixed probe geometry, changing the nonbonded probe--host affinity can alter the local mobility environments sampled by the probe. Probe size and probe--host affinity therefore provide distinct, although not necessarily independent, contributions to the dynamics reported by a molecular probe. Different modes of probe--host coupling can also lead to distinct dynamical behavior. Covalently tethered probes, for example, can remain faithful reporters of host segmental dynamics even under persistent mechanical attachment\cite{Vallee2010b,Choi2019}. This situation differs from the present model, in which increasing \(\varepsilon_{pd}\) strengthens a nonbonded affinity without permanently attaching the probe to a particular polymer segment; the probe remains translationally mobile even at the highest interaction strength examined. The present interaction dependence is therefore better viewed as a change in the local environments sampled by a mobile probe than as a simple increase in mechanical constraint.

\begin{figure}
    \centering
    \includegraphics[width=0.5\linewidth]{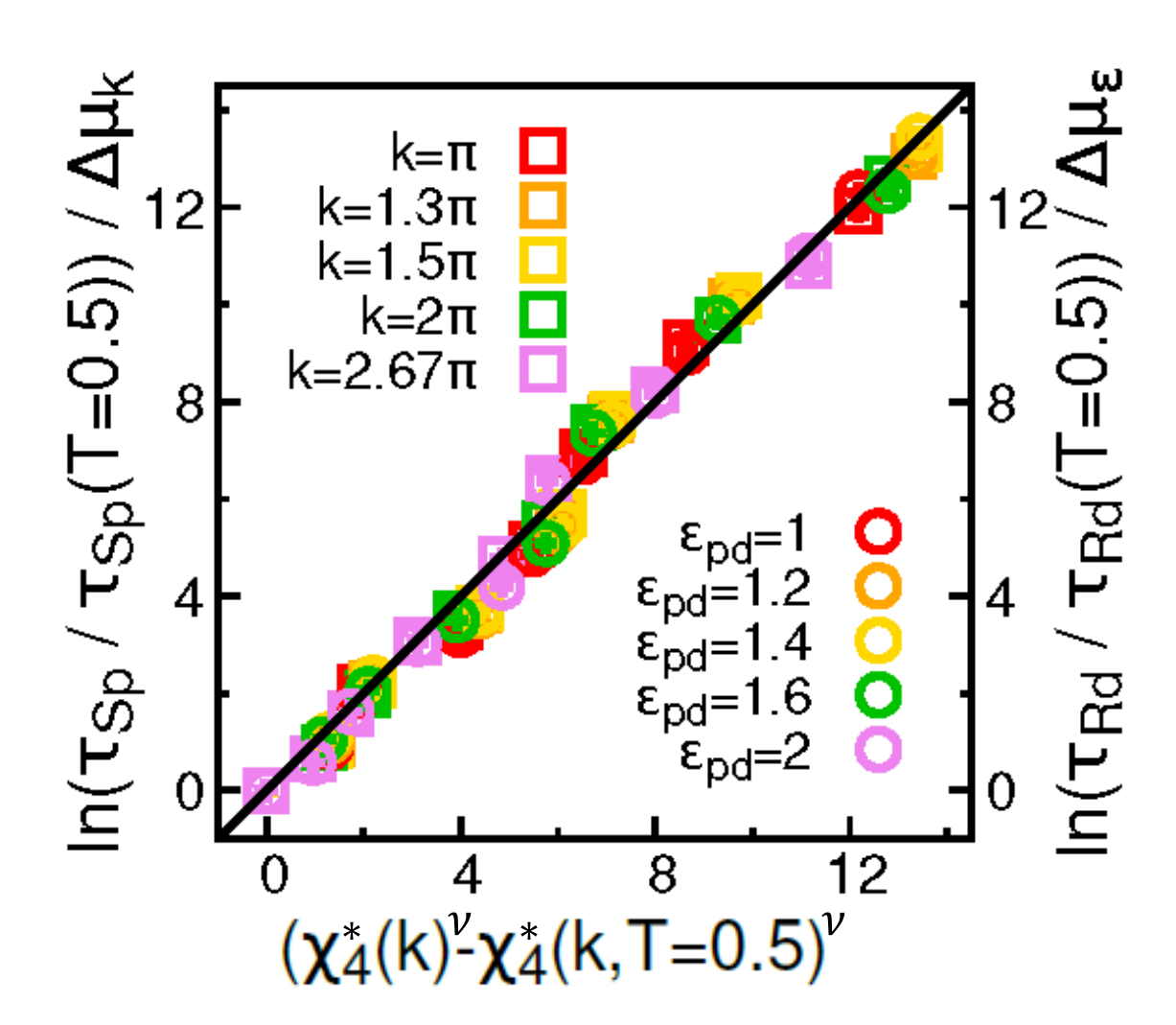}
    \caption{Phenomenological scaling of host and probe logarithmic relaxation-time shifts with dynamic heterogeneity. The normalized logarithmic relaxation-time shifts of the bulk polymer, $\ln[\tau_{Sp}(k,T)/\tau_{Sp}(k,T=0.5)]/\Delta\mu_k$, and the probe dimer, $\ln[\tau_{Rd}(\varepsilon_{pd},T)/\tau_{Rd}(\varepsilon_{pd},T=0.5)]/\Delta\mu_{\varepsilon}$, are plotted against $[\chi_4^{*}(k,T)^{\nu}-\chi_4^{*}(k,T=0.5)^{\nu}]/T$, where $\chi_4^{*}(k,T)$ is the peak value of the $k$-dependent dynamic susceptibility $\chi_4(k,t)$. For each selected probe interaction strength, the corresponding effective matching wavevector identified from Figure~\ref{fig3} is used. The peak values $\chi_4^{*}(k,T)$ were determined by quadratic fitting near the local maxima of $\chi_4(k,t)$. The exponent is fixed at $\nu=0.556$, and the solid black line denotes $y=x$.}
    \label{fig7}
\end{figure}

The real-space results above show that changing $\varepsilon_{pd}$ modifies the local dynamical environments sampled by the probe. We next examine whether the temperature evolution of the probe and host relaxation can be represented within a common dynamic-heterogeneity-based scaling framework. We use the $k$-dependent dynamic susceptibility $\chi_4(k,t)$ (Figure S19 in the Supporting Information) and its peak value $\chi_4^{*}(k,T)$ as a phenomenological, wavevector-dependent measure of dynamic heterogeneity.

Motivated by Adam--Gibbs-type connections between relaxation and cooperative dynamics, we use \([\chi_4^{*}(k,T)]^{\nu}\) as a phenomenological scaling variable for describing the temperature evolution of dynamic heterogeneity. Specifically, we test the relaxation data using
\[
\frac{\chi_4^{*}(k,T)^{\nu}-\chi_4^{*}(k,T_{\mathrm{ref}})^{\nu}}{T},
\]
where $T_{\mathrm{ref}}=0.5$ and $\nu$ controls how the susceptibility amplitude enters the scaling representation.

Figure~\ref{fig7} shows the normalized logarithmic relaxation-time shifts of the bulk polymer,
\[
\ln[\tau_{Sp}(k,T)/\tau_{Sp}(k,T=0.5)]/\Delta\mu_k,
\]
and the probe dimers,
\[
\ln[\tau_{Rd}(\varepsilon_{pd},T)/\tau_{Rd}(\varepsilon_{pd},T=0.5)]/\Delta\mu_{\varepsilon},
\]
as a function of the susceptibility-based scaling variable. With this normalization, the host relaxation data at different wavevectors and the probe relaxation data at different interaction strengths approximately collapse onto the same scaling relation.

The scaling procedure was performed in two steps. We first determined the exponent $\nu$ using only the bulk-polymer relaxation data. For each trial value of $\nu$, the polymer data at different wavevectors were fitted by varying the normalization factors $\Delta\mu_k$, and the weighted residual sum was used to identify an optimal value, $\nu=0.556$ (Figure S20 in the Supporting Information). This exponent is used here only as a phenomenological scaling parameter for comparing the temperature evolution of relaxation at different wavevectors.

Using this host-derived exponent, we then analyzed the probe relaxation data by varying only the normalization factors $\Delta\mu_{\varepsilon}$. For each $\varepsilon_{pd}$, the host susceptibility is evaluated at the corresponding effective matching wavevector identified in Figure~\ref{fig3}; the probe data are not used to determine $\nu$. The resulting probe data lie approximately on the same scaling line as the host data, and the fitted values of $\Delta\mu_{\varepsilon}$ are close to the corresponding $\Delta\mu_k$ values obtained from the bulk-polymer relaxation (Table~S2 in the Supporting Information). This phenomenological collapse shows that the temperature evolution of the probe relaxation can be represented consistently using the host-derived dynamic-heterogeneity scaling obtained from the wavevector-dependent polymer relaxation.

Together with the real-space analyses in Figures~\ref{fig5} and~\ref{fig6}, this scaling provides complementary context for the interaction-dependent probe--host relaxation behavior. The displacement distributions and local mobility statistics show that changing $\varepsilon_{pd}$ modifies the dynamical environments explored by the probe, while the susceptibility-based representation relates the temperature evolution of probe relaxation to the wavevector-resolved host dynamics. These results provide a microscopic context for the systematic interaction dependence of the probe--host scaling.

\section{Conclusion}

In this study, we investigated how molecular probes report the dynamics of a spatially heterogeneous supercooled polymer matrix. Our results show that probe relaxation does not simply reproduce the intrinsic relaxation of the host, even in the dilute-probe limit where the bulk matrix remains essentially unperturbed. Instead, the probe--host interaction strength influences the probe relaxation behavior and the local dynamical environments sampled by the probe. Lower-affinity probes exhibit stronger non-Gaussian translational motion and broader displacement distributions, whereas higher-affinity probes remain more localized and are statistically less exposed to the most dynamically active host environments. These differences provide a microscopic picture of how probe--host affinity influences the dynamics reported by molecular probes.

The interaction-dependent probe--host correspondence is also reflected in comparisons with wavevector-dependent host structural relaxation. For each interaction strength, an effective matching wavevector can be identified at which the host structural relaxation exhibits a temperature dependence most similar to that of the probe relaxation. We use this matching wavevector as a phenomenological measure of the correspondence between probe and host dynamics. The accompanying analysis based on the \(k\)-dependent dynamic susceptibility further shows that the temperature evolution of the probe relaxation can be represented consistently within the same host-derived dynamic-heterogeneity framework. Together with the real-space analysis, these results indicate that probe--host affinity is associated with systematic changes in the local dynamical environments sampled by the probe and in its correspondence with host relaxation.

The present coarse-grained dimer model is intentionally minimal and does not represent the detailed architecture or chemical interactions of specific experimental probes. Additional calculations with different probe sizes show that the interaction-dependent probe--host scaling is not restricted to a single probe dimension, although its quantitative magnitude depends on probe size. More generally, probe size, shape, and chemical specificity are all expected to influence the correspondence between probe and host dynamics. The present results identify probe--host affinity as one factor that can influence this correspondence. Extending this approach to more realistic probe architectures and chemically specific interactions may further clarify how molecular probes report local glassy dynamics in complex and heterogeneous polymeric materials.

\bibliography{dimer}

\section*{Acknowledgments}
This work was supported by the research grant of Jeju National University in 2025.

\section*{Author Declarations}
\subsection*{Conflict of Interest}
The authors have no conflicts to disclose.
\subsection*{Author Contributions}
S.K. performed the simulations, curated and analyzed the data, and contributed to validation and visualization. T.K. conceived and supervised the project, secured funding, and administered the project. Both authors discussed the results, wrote the manuscript, reviewed and edited the final version, and approved the final manuscript.

\section*{Data Availability}
The data that support the findings of this study are in the article and available from the corresponding author upon reasonable request.

\section*{Supporting Information}

Additional results and analyses of host and probe relaxation, VFT fitting and fragility, probe--host relaxation scaling, probe-size effects, wavevector-dependent structural relaxation, mean-squared displacement, dynamic heterogeneity, and fitted scaling parameters.

\clearpage
\setcounter{page}{1}
\setcounter{figure}{0}
\setcounter{table}{0}
\renewcommand{\thefigure}{S\arabic{figure}}
\renewcommand{\thetable}{S\arabic{table}}

\begin{center}
{\Large\bfseries Interaction-Dependent Molecular Probing of Heterogeneous Glassy Dynamics in Polymer Melts\par}
\vspace{1.2em}
{\large Suyeon Kim$^{1}$ and Taejin Kwon$^{2,*}$\par}
\vspace{0.8em}
{\normalsize $^{1}$Department of Chemical Education, Jeju National University, Jeju 63243, Republic of Korea\par}
{\normalsize $^{2}$Department of Chemistry and Cosmetics, Jeju National University, Jeju 63243, Republic of Korea\par}
\vspace{0.6em}
{\normalsize $^{*}$tjkwon@jejunu.ac.kr\par}
\end{center}
\clearpage

\begin{figure}
    \centering
    \includegraphics[width=1\linewidth]{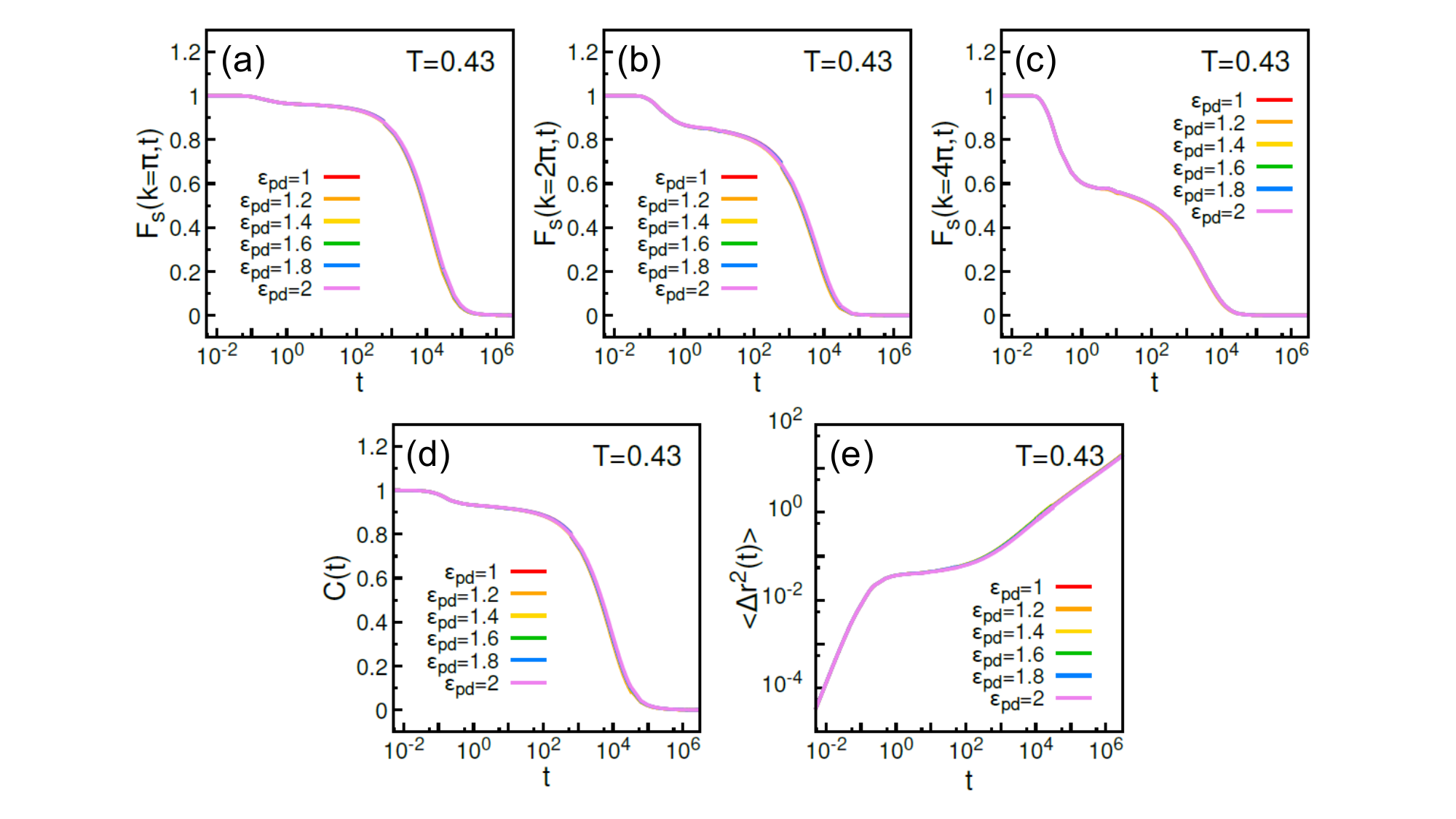}
\caption{Dynamics of the host polymer matrix with probe dimers at $T=0.43$. (a--c) Self-intermediate scattering functions ($F_s(k,t)$) of the host polymer monomers at (a) $k=\pi/\sigma$, (b) $k=2\pi/\sigma$, and (c) $k=4\pi/\sigma$ for different probe--host interaction strengths $\varepsilon_{pd}$. (d) Bond orientational time correlation function ($C(t)$) of the host polymer matrix for different $\varepsilon_{pd}$. The correlation function $C(t)$ is calculated from the bond vectors between adjacent bonded monomers along the polymer chain. (e) Mean-squared displacement $\langle \Delta r^2(t)\rangle$ of the monomers for different $\varepsilon_{pd}$.
}
\label{figS1}
\end{figure}

\clearpage
\begin{figure}
    \centering
    \includegraphics[width=1\linewidth]{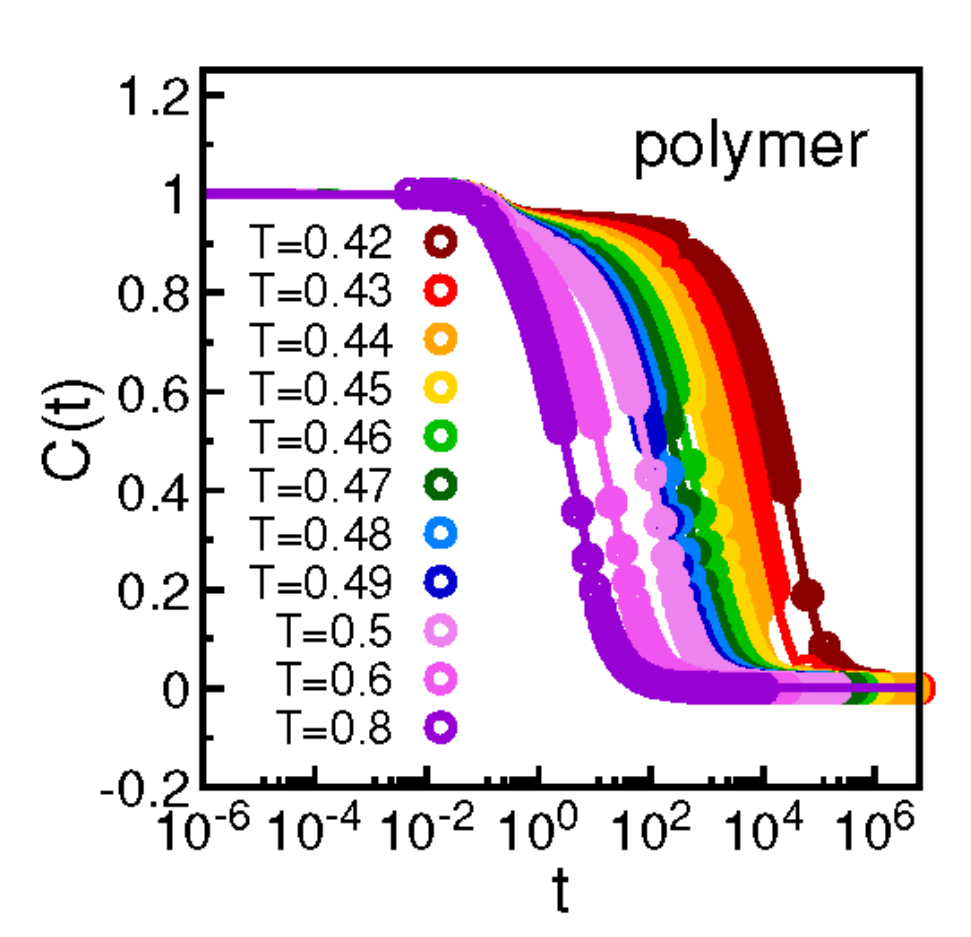}
\caption{Bond orientational time correlation function ($C(t)$) of the bulk polymer. Markers represent simulation results, and solid lines denote fits using a combination of exponential and stretched exponential functions.
}
\label{figS2}
\end{figure}

\clearpage

\begin{figure}
    \centering
    \includegraphics[width=1\linewidth]{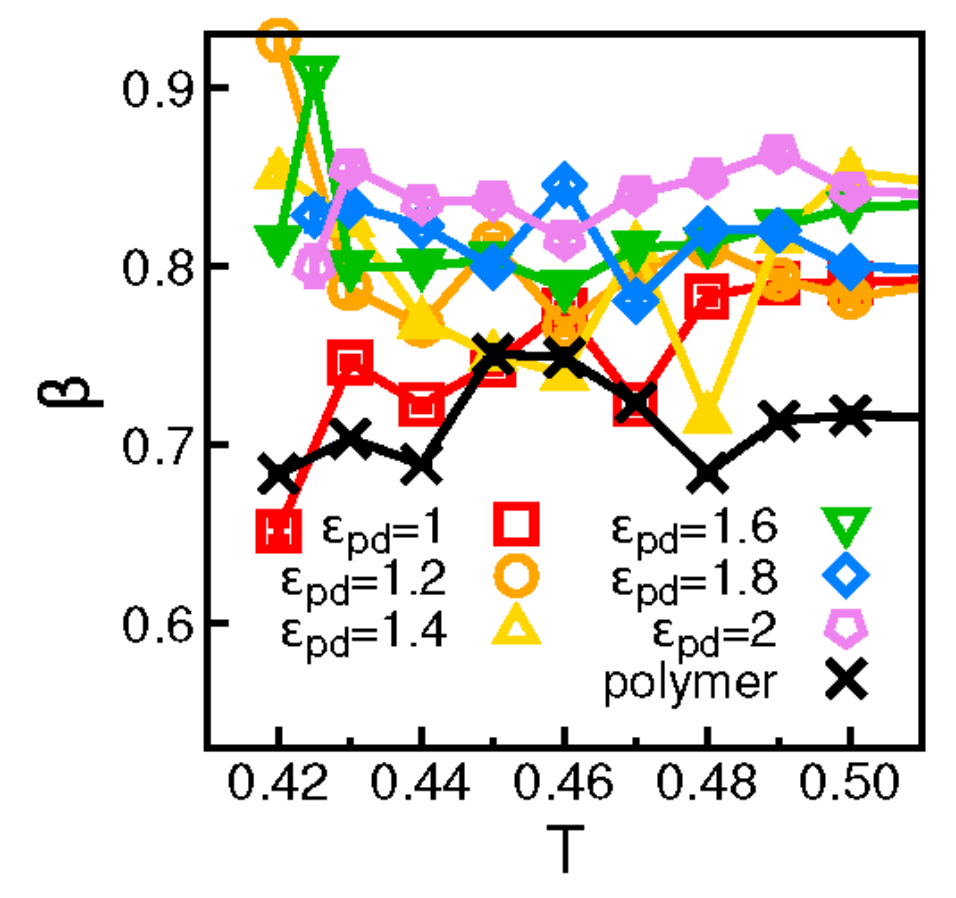}
    \caption{Temperature dependence of the stretching exponent $\beta$. The stretching exponent $\beta$ obtained from relaxation function fits is shown as a function of temperature for the probe dimers at different interaction strengths $\varepsilon_{pd}$ and for the polymer.
}
    \label{figS3}
\end{figure}

\clearpage
\begin{figure}
    \centering
    \includegraphics[width=1\linewidth]{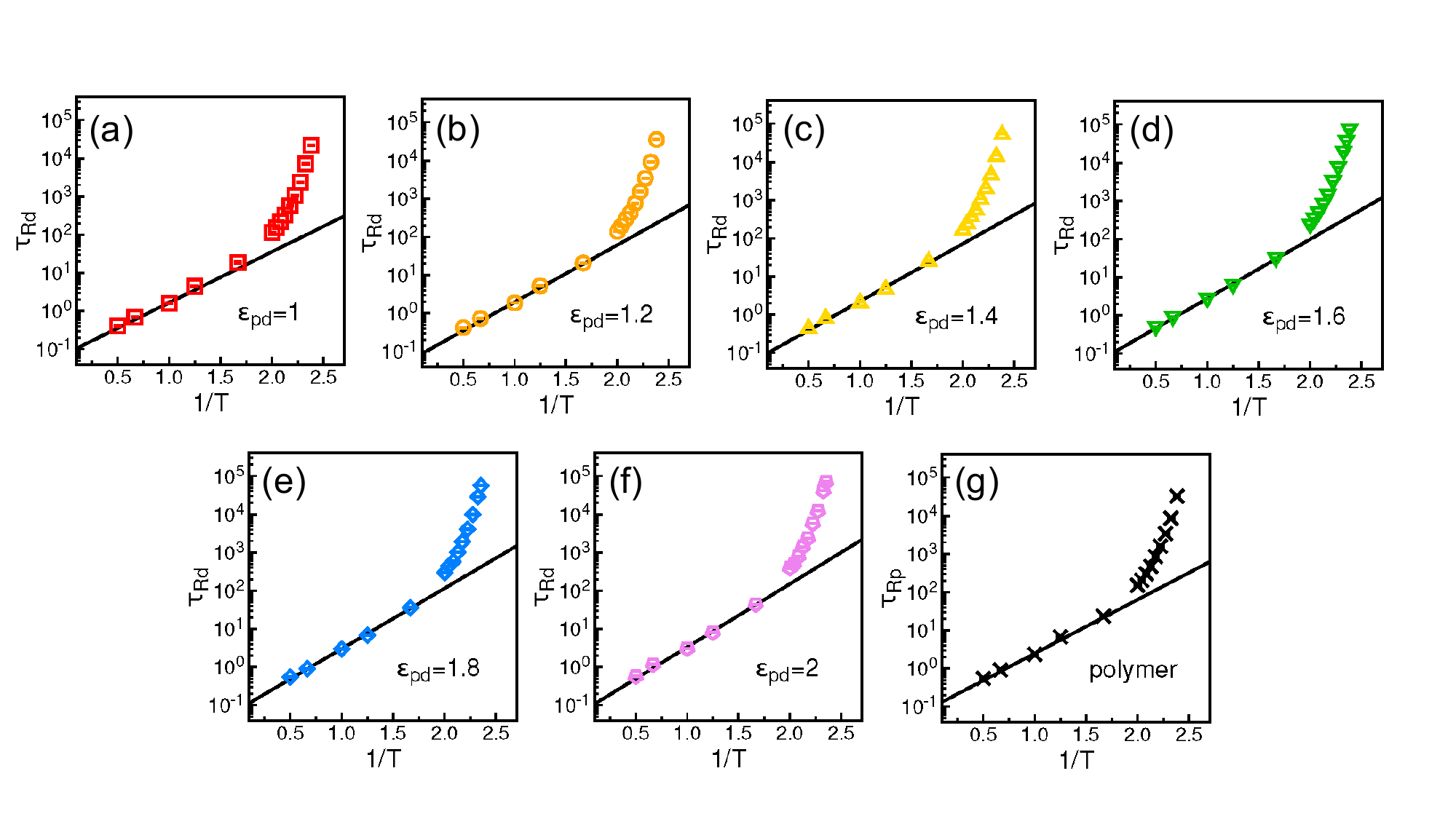}
    \caption{Temperature dependence of rotational relaxation times. (a--f) The rotational relaxation time of the probe dimers ($\tau_{Rd}$) as a function of inverse temperature $1/T$ for (a) $\varepsilon_{pd}=1$, (b) $\varepsilon_{pd}=1.2$, (c) $\varepsilon_{pd}=1.4$, (d) $\varepsilon_{pd}=1.6$, (e) $\varepsilon_{pd}=1.8$, and (f) $\varepsilon_{pd}=2$. (g) Rotational relaxation time of the bulk polymer ($\tau_{Rp}$) as a function of inverse temperature. Markers represent simulation results, and solid lines represent Arrhenius fits to the high temperature regime.
}
    \label{figS4}
\end{figure}

\begin{table*}[htbp]
\centering
\caption{VFT fitting parameters obtained using different fitting constraints.
The VFT relation is given by $\tau=\tau_0\exp[B/(T-T_0)]$.
Standard errors (SE) are reported for fitted parameters.
For fixed parameters, the corresponding SE is denoted by ``--''.
Parameter correlations are reported for fitting protocols in which both $B$ and $T_0$
are simultaneously optimized.}
\label{tab:S1}
\scriptsize
\setlength{\tabcolsep}{4.5pt}
\renewcommand{\arraystretch}{1.15}

\begin{tabular}{lccccccc}
\hline
\hline
 & Polymer & $\varepsilon_{pd}=1$ & $\varepsilon_{pd}=1.2$ &
 $\varepsilon_{pd}=1.4$ & $\varepsilon_{pd}=1.6$ &
 $\varepsilon_{pd}=1.8$ & $\varepsilon_{pd}=2$ \\
\hline

\multicolumn{8}{l}{\textbf{(I) Fixed $B_p$ and $T_{0,p}$; fitted $\tau_0$}} \\

$\tau_0$ &
$7.49\times10^{-1}$ &
$5.49\times10^{-1}$ &
$7.13\times10^{-1}$ &
$1.03\times10^{0}$ &
$1.31\times10^{0}$ &
$1.77\times10^{0}$ &
$2.22\times10^{0}$ \\

$B$ &
$8.40\times10^{-1}$ &
$8.40\times10^{-1}$ &
$8.40\times10^{-1}$ &
$8.40\times10^{-1}$ &
$8.40\times10^{-1}$ &
$8.40\times10^{-1}$ &
$8.40\times10^{-1}$ \\

$T_0$ &
$3.40\times10^{-1}$ &
$3.40\times10^{-1}$ &
$3.40\times10^{-1}$ &
$3.40\times10^{-1}$ &
$3.40\times10^{-1}$ &
$3.40\times10^{-1}$ &
$3.40\times10^{-1}$ \\

SE($\tau_0$) &
$1.858\times10^{-1}$ &
$1.46\times10^{-2}$ &
$1.25\times10^{-2}$ &
$4.69\times10^{-2}$ &
$5.52\times10^{-2}$ &
$1.13\times10^{-1}$ &
$1.43\times10^{-1}$ \\

SE($B$) &
$5.68\times10^{-2}$ &
-- & -- & -- & -- & -- & -- \\

SE($T_0$) &
$3.70\times10^{-3}$ &
-- & -- & -- & -- & -- & -- \\

\hline

\multicolumn{8}{l}{\textbf{(II) Fixed $\tau_{0,p}$; fitted $B$ and $T_0$}} \\

$\tau_0$ &
$7.49\times10^{-1}$ &
$7.49\times10^{-1}$ &
$7.49\times10^{-1}$ &
$7.49\times10^{-1}$ &
$7.49\times10^{-1}$ &
$7.49\times10^{-1}$ &
$7.49\times10^{-1}$ \\

$B$ &
$8.40\times10^{-1}$ &
$7.53\times10^{-1}$ &
$8.06\times10^{-1}$ &
$8.38\times10^{-1}$ &
$8.99\times10^{-1}$ &
$9.31\times10^{-1}$ &
$9.85\times10^{-1}$ \\

$T_0$ &
$3.40\times10^{-1}$ &
$3.47\times10^{-1}$ &
$3.44\times10^{-1}$ &
$3.45\times10^{-1}$ &
$3.42\times10^{-1}$ &
$3.42\times10^{-1}$ &
$3.39\times10^{-1}$ \\

SE($\tau_0$) &
$1.858\times10^{-1}$ &
-- & -- & -- & -- & -- & -- \\

SE($B$) &
$5.68\times10^{-2}$ &
$1.14\times10^{-2}$ &
$1.01\times10^{-2}$ &
$9.85\times10^{-3}$ &
$9.79\times10^{-3}$ &
$1.29\times10^{-2}$ &
$1.76\times10^{-2}$ \\

SE($T_0$) &
$3.70\times10^{-3}$ &
$1.38\times10^{-3}$ &
$1.38\times10^{-3}$ &
$1.12\times10^{-3}$ &
$1.18\times10^{-3}$ &
$1.44\times10^{-3}$ &
$2.01\times10^{-3}$ \\

$\rho(B,T_0)$ &
-- &
$-0.981$ &
$-0.986$ &
$-0.986$ &
$-0.981$ &
$-0.987$ &
$-0.985$ \\

\hline

\multicolumn{8}{l}{\textbf{(III) Fixed probe-specific $\tau_{0,d}^{\mathrm{shift}}$; fitted $B$ and $T_0$}} \\

$\tau_0$ &
$7.49346\times10^{-1}$ &
$5.48555\times10^{-1}$ &
$7.12994\times10^{-1}$ &
$1.03216\times10^{0}$ &
$1.31312\times10^{0}$ &
$1.77487\times10^{0}$ &
$2.22082\times10^{0}$ \\

$B$ &
$8.40410\times10^{-1}$ &
$8.18416\times10^{-1}$ &
$8.17547\times10^{-1}$ &
$7.70525\times10^{-1}$ &
$7.76593\times10^{-1}$ &
$7.43014\times10^{-1}$ &
$7.45280\times10^{-1}$ \\

$T_0$ &
$3.40150\times10^{-1}$ &
$3.42939\times10^{-1}$ &
$3.43438\times10^{-1}$ &
$3.49237\times10^{-1}$ &
$3.49436\times10^{-1}$ &
$3.53574\times10^{-1}$ &
$3.54149\times10^{-1}$ \\

SE($\tau_0$) &
$1.858\times10^{-1}$ &
-- & -- & -- & -- & -- & -- \\

SE($B$) &
$5.68\times10^{-2}$ &
$1.375\times10^{-2}$ &
$1.062\times10^{-2}$ &
$7.244\times10^{-3}$ &
$9.706\times10^{-3}$ &
$8.645\times10^{-3}$ &
$1.255\times10^{-2}$ \\

SE($T_0$) &
$3.70\times10^{-3}$ &
$1.616\times10^{-3}$ &
$1.447\times10^{-3}$ &
$8.467\times10^{-4}$ &
$1.263\times10^{-3}$ &
$1.051\times10^{-3}$ &
$1.610\times10^{-3}$ \\

$\rho(B,T_0)$ &
-- &
$-0.983$ &
$-0.986$ &
$-0.984$ &
$-0.978$ &
$-0.984$ &
$-0.971$ \\

\hline

\multicolumn{8}{l}{\textbf{(IV) Fixed $\tau_{0,p}$ and $B_p$; fitted $T_0$}} \\

$\tau_0$ &
$7.49346\times10^{-1}$ &
$7.49346\times10^{-1}$ &
$7.49346\times10^{-1}$ &
$7.49346\times10^{-1}$ &
$7.49346\times10^{-1}$ &
$7.49346\times10^{-1}$ &
$7.49346\times10^{-1}$ \\

$B$ &
$8.40410\times10^{-1}$ &
$8.40410\times10^{-1}$ &
$8.40410\times10^{-1}$ &
$8.40410\times10^{-1}$ &
$8.40410\times10^{-1}$ &
$8.40410\times10^{-1}$ &
$8.40410\times10^{-1}$ \\

$T_0$ &
$3.40150\times10^{-1}$ &
$3.35951\times10^{-1}$ &
$3.39448\times10^{-1}$ &
$3.44702\times10^{-1}$ &
$3.48631\times10^{-1}$ &
$3.51506\times10^{-1}$ &
$3.54733\times10^{-1}$ \\

SE($\tau_0$) &
$1.858\times10^{-1}$ &
-- & -- & -- & -- & -- & -- \\

SE($B$) &
$5.68\times10^{-2}$ &
-- & -- & -- & -- & -- & -- \\

SE($T_0$) &
$3.70\times10^{-3}$ &
$8.517\times10^{-4}$ &
$3.527\times10^{-4}$ &
$1.774\times10^{-4}$ &
$4.926\times10^{-4}$ &
$5.329\times10^{-4}$ &
$9.138\times10^{-4}$ \\

\hline
\hline
\end{tabular}

\vspace{4pt}

\begin{minipage}{0.98\textwidth}
\footnotesize
The reference polymer VFT parameters were obtained by simultaneously fitting
$\tau_0$, $B$, and $T_0$. The corresponding parameter correlation coefficients are
$\rho(B,T_0)=-0.998$,
$\rho(\tau_0,B)=-0.996$, and
$\rho(\tau_0,T_0)=0.988$.
\end{minipage}

\end{table*}

\clearpage
\begin{figure}
    \centering
    \includegraphics[width=1\linewidth]{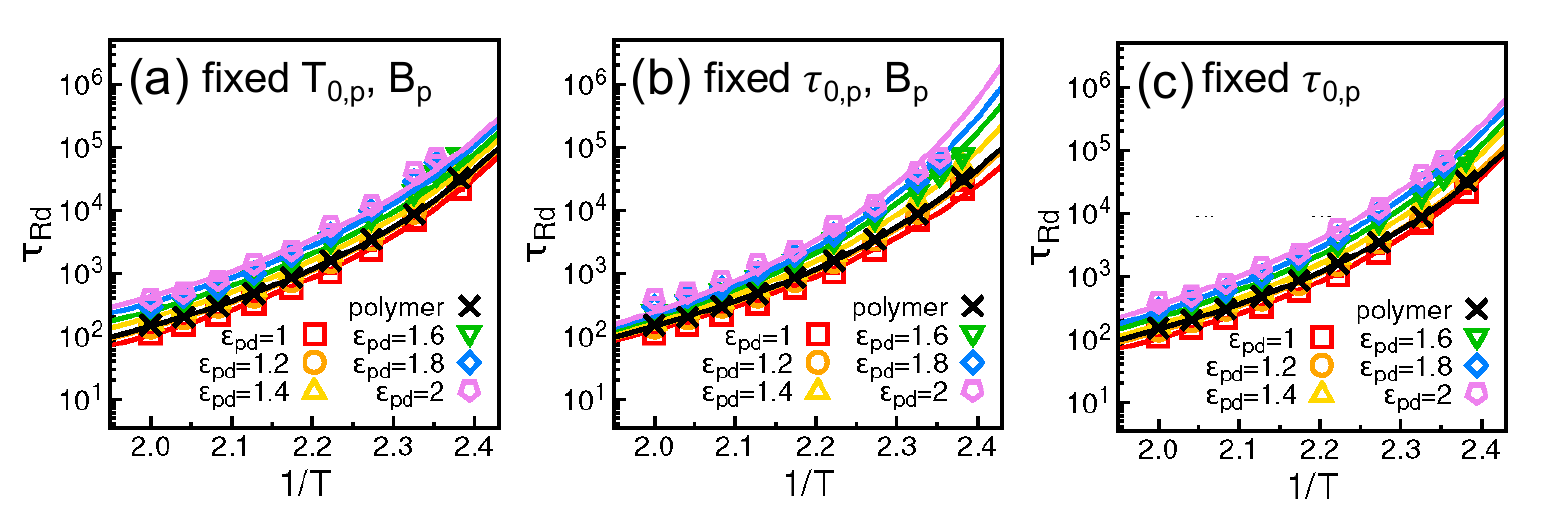}
\caption{Alternative constrained VFT fits to the probe rotational relaxation times. 
Probe rotational relaxation times, $\tau_{Rd}$, are shown as a function of inverse temperature for $\varepsilon_{pd}=1\varepsilon$--$2\varepsilon$. Black crosses represent the polymer bond-orientational relaxation time, $\tau_{Rp}$, and colored open symbols represent the probe relaxation times. Solid lines denote VFT fits, $\tau=\tau_0^{\mathrm{VFT}}\exp[B/(T-T_0)]$, obtained using different parameter constraints: (a) $B_p$ and $T_{0,p}$ are fixed to the polymer values and only $\tau_0$ is fitted; (b) $\tau_{0,p}$ and $B_p$ are fixed and only $T_0$ is fitted; and (c) $\tau_{0,p}$ is fixed while $B$ and $T_0$ are fitted. The complementary fit in which the probe-specific prefactor $\tau_{0,d}^{\mathrm{shift}}$ obtained from the vertical-shift analysis is fixed while $B_d$ and $T_{0,d}$ are fitted is shown in Figure~1(d).}
    \label{figS5}
\end{figure}

\clearpage
\begin{figure}
    \centering
    \includegraphics[width=1\linewidth]{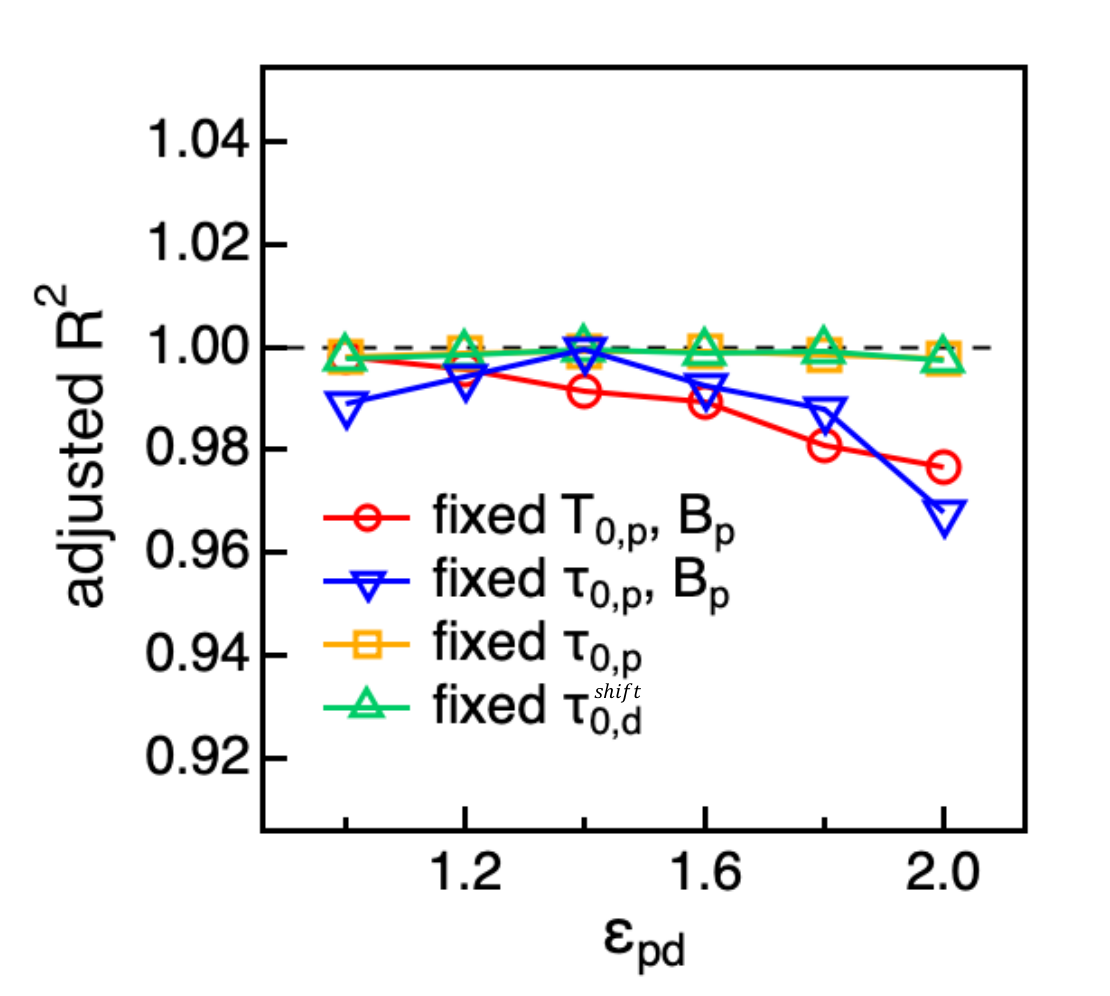}
\caption{Goodness of fit for the different constrained VFT descriptions of the probe rotational relaxation. Adjusted \(R^2\) values are shown as a function of \(\varepsilon_{pd}\) for four fitting protocols: (1) \(B_p\) and \(T_{0,p}\) fixed to the polymer values while \(\tau_0\) is fitted; (2) \(\tau_{0,p}\) and \(B_p\) fixed while \(T_0\) is fitted; (3) \(\tau_{0,p}\) fixed while \(B\) and \(T_0\) are fitted; and (4) the probe-specific prefactor \(\tau_{0,d}^{\mathrm{shift}}\), obtained from the vertical-shift analysis, fixed while \(B_d\) and \(T_{0,d}\) are fitted. The coefficient of determination was calculated in log-relaxation-time space as \(R^2=1-\sum_i[\log_{10}\tau_i^{\mathrm{sim}}-\log_{10}\tau_i^{\mathrm{VFT}}]^2/\sum_i[\log_{10}\tau_i^{\mathrm{sim}}-\overline{\log_{10}\tau^{\mathrm{sim}}}]^2\), and was adjusted for the number of fitted parameters according to \(R_{\mathrm{adj}}^2=1-(1-R^2)(n-1)/(n-p-1)\), where \(n=9\) is the number of temperature points and \(p\) is the number of freely fitted VFT parameters. Thus, \(p=1\) was used for protocols (1) and (2), whereas \(p=2\) was used for protocols (3) and (4). The horizontal dashed line denotes \(R_{\mathrm{adj}}^2=1\).}
\label{figS6}
\end{figure}

\clearpage
\begin{figure}
    \centering
    \includegraphics[width=1\linewidth]{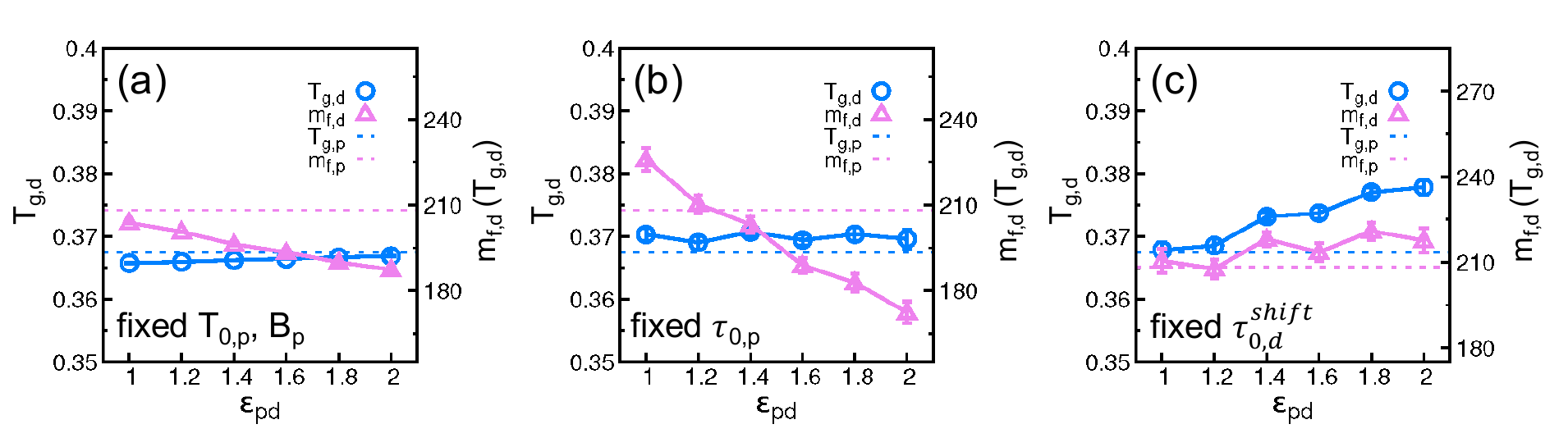}
\caption{Dependence of the probe-derived glass-transition temperature and apparent fragility on the VFT fitting constraint. 
The probe glass-transition temperature, $T_{g,d}$ (blue circles, left axis), and apparent fragility evaluated at the corresponding probe-derived glass-transition temperature, $m_{f,d}(T_{g,d})$ (magenta triangles, right axis), are shown as functions of $\varepsilon_{pd}$. The operational probe glass-transition temperature is defined by $\tau_{Rd}(T_{g,d})=10^{14}\tau$, and the apparent fragility is evaluated from the VFT parameters as $m_f=BT_g/[\ln(10)(T_g-T_0)^2]$. VFT fits were performed with (a) $B_p$ and $T_{0,p}$ fixed to the polymer values while $\tau_0$ was fitted, (b) $\tau_{0,p}$ fixed while $B$ and $T_0$ were fitted, and (c) the probe-specific prefactor $\tau_{0,d}^{\mathrm{shift}}$, obtained from the vertical-shift analysis, fixed while $B_d$ and $T_{0,d}$ were fitted. Blue and magenta dashed lines denote the corresponding polymer values, $T_{g,p}$ and $m_{f,p}$, respectively. Error bars were obtained by first-order propagation of the fitted VFT-parameter uncertainties, including the covariance between $B$ and $T_0$ when both parameters were fitted; for panel (c), the propagated uncertainty is conditional on the fixed $\tau_{0,d}^{\mathrm{shift}}$.}
    \label{figS7}
\end{figure}

\clearpage
\begin{figure}
    \centering
    \includegraphics[width=1\linewidth]{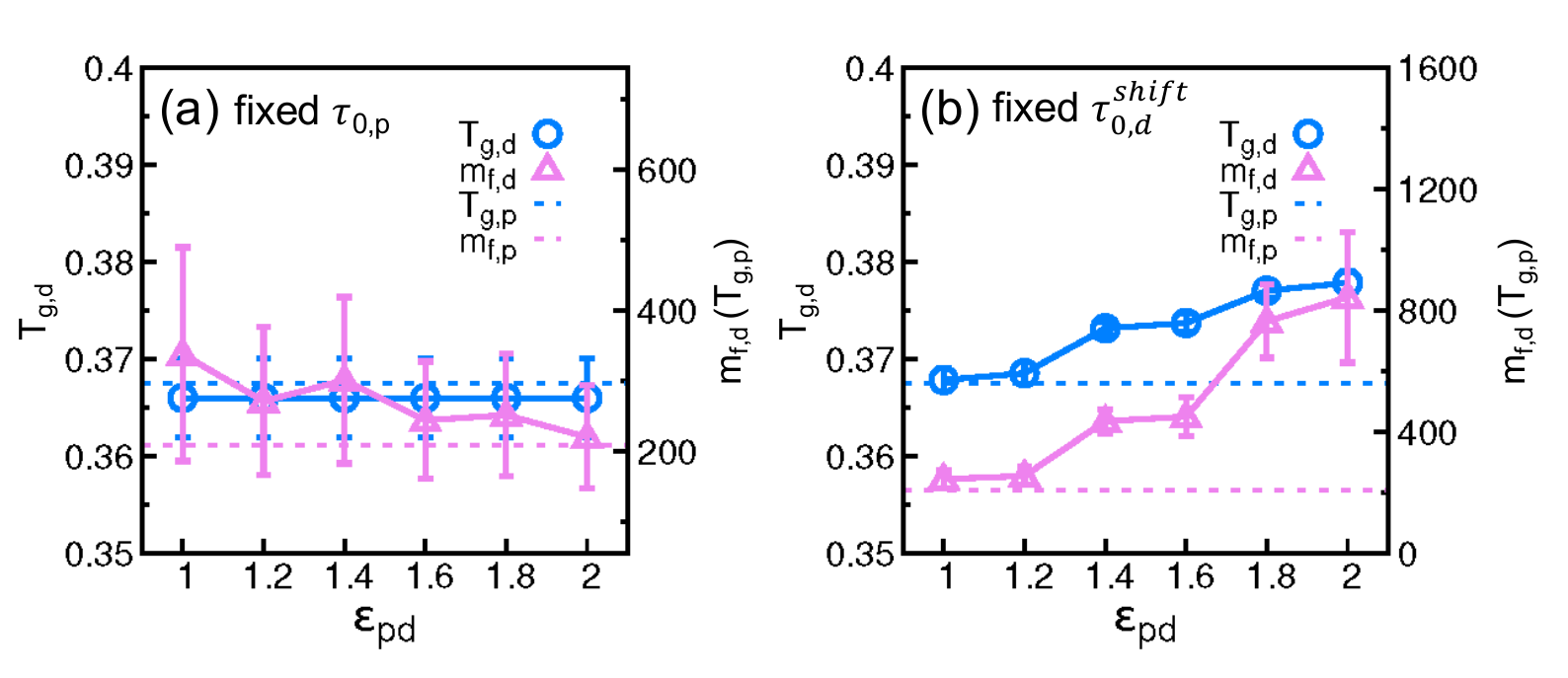}
\caption{Sensitivity of the apparent probe fragility to the temperature at which the VFT slope is evaluated. 
The probe-derived glass-transition temperature, $T_{g,d}$ (blue circles, left axis), and apparent probe fragility evaluated at the common polymer glass-transition temperature, $m_{f,d}(T_{g,p})$ (magenta triangles, right axis), are shown as functions of $\varepsilon_{pd}$. VFT fits were performed with (a) $\tau_{0,p}$ fixed while $B$ and $T_0$ were fitted and (b) the probe-specific prefactor $\tau_{0,d}^{\mathrm{shift}}$ obtained from the vertical-shift analysis fixed while $B_d$ and $T_{0,d}$ were fitted. The fragility is evaluated using the probe VFT parameters at the common polymer temperature $T_{g,p}$ rather than at the corresponding $T_{g,d}$. Blue and magenta dashed lines denote the polymer $T_{g,p}$ and $m_{f,p}$, respectively. Error bars were obtained by first-order propagation of the fitted $B$ and $T_0$ uncertainties, including their covariance; for panel (b), the reported uncertainty is conditional on the fixed $\tau_{0,d}^{\mathrm{shift}}$.}
    \label{figS8}
\end{figure}

\clearpage
\begin{figure}
    \centering
    \includegraphics[width=1\linewidth]{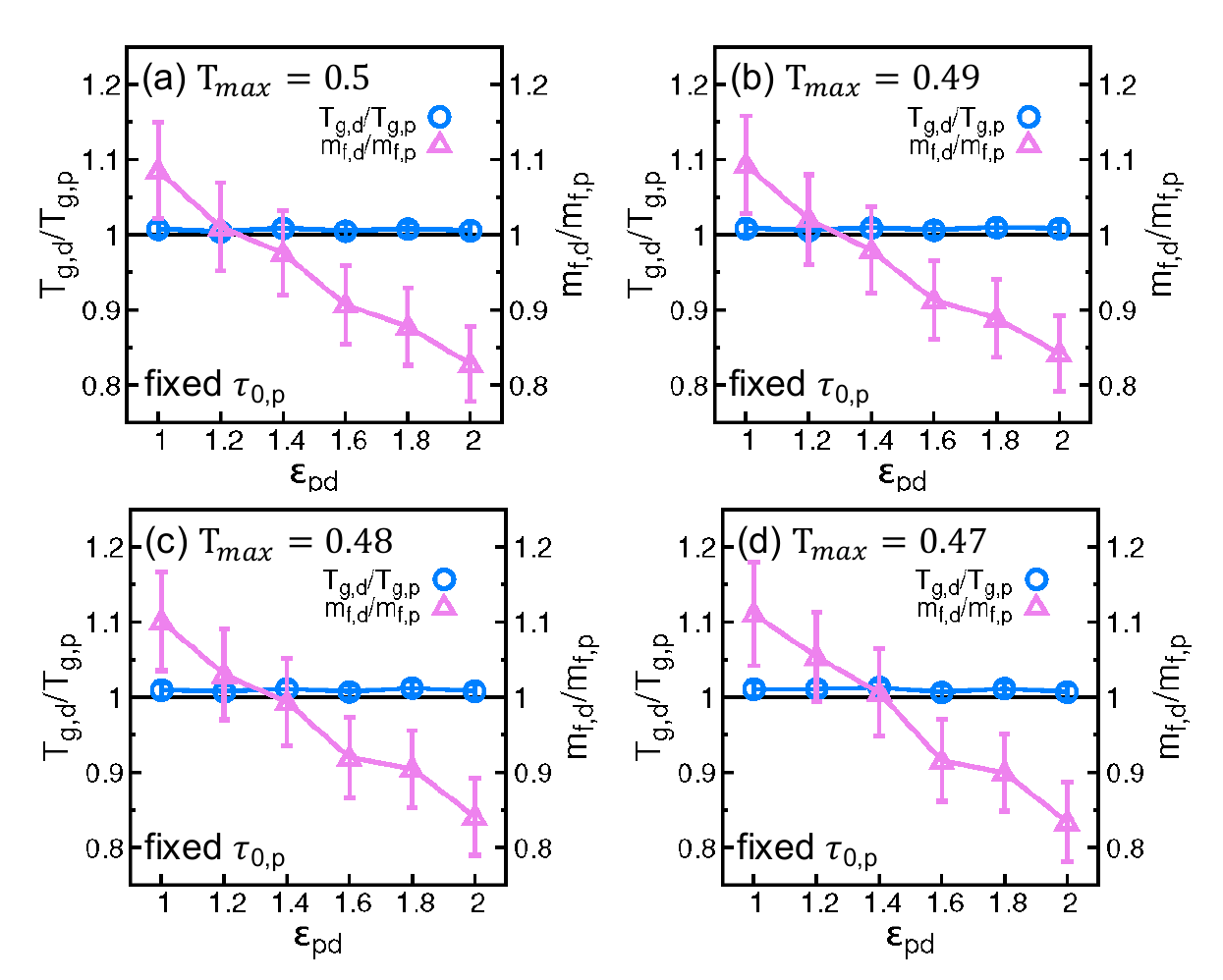}
\caption{Sensitivity of the probe-derived glass-transition temperature and apparent fragility to the VFT fitting window for common-prefactor fits.
The normalized probe glass-transition temperature, $T_{g,d}/T_{g,p}$ (blue circles), and normalized apparent fragility, $m_{f,d}/m_{f,p}$ (magenta triangles), are shown as functions of $\varepsilon_{pd}$ for VFT fits performed using upper temperature limits of (a) $T_{\max}=0.50$, (b) 0.49, (c) 0.48, and (d) 0.47.
In all panels, $\tau_{0,p}$ is fixed to the polymer value while $B$ and $T_0$ are fitted.
All available low-temperature data are retained, with the minimum simulated temperature ranging from $T=0.42$ to 0.425 depending on $\varepsilon_{pd}$.
$T_{g,d}$ is defined by $\tau_{Rd}(T_{g,d})=10^{14}\tau$, and $m_{f,d}$ is evaluated at the corresponding $T_{g,d}$.
Horizontal lines at unity denote the polymer reference values.
Error bars were obtained by first-order propagation of the fitted $B$ and $T_0$ uncertainties, including their covariance.}
\label{figS9}
\end{figure}

\clearpage
\begin{figure}
    \centering
    \includegraphics[width=1\linewidth]{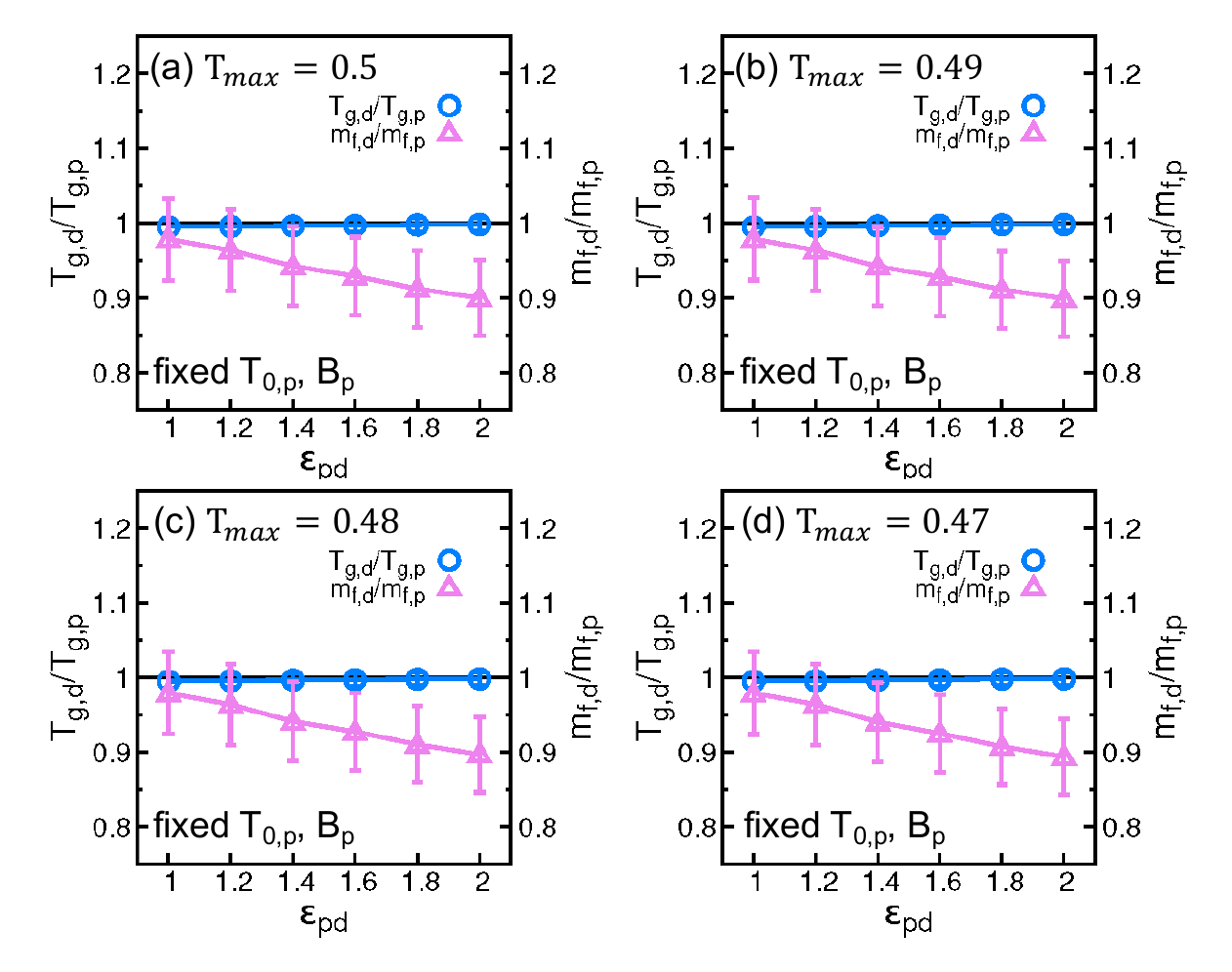}
\caption{Sensitivity of the probe-derived glass-transition temperature and apparent fragility to the VFT fitting window for vertical-shift-constrained fits. 
The normalized probe glass-transition temperature, $T_{g,d}/T_{g,p}$ (blue circles), and normalized apparent fragility, $m_{f,d}/m_{f,p}$ (magenta triangles), are shown as functions of $\varepsilon_{pd}$ for VFT fits performed using upper temperature limits of (a) $T_{\max}=0.50$, (b) 0.49, (c) 0.48, and (d) 0.47. In all panels, $B_p$ and $T_{0,p}$ are fixed to the polymer values and only the probe prefactor $\tau_0$ is fitted. All available low-temperature data are retained, with the minimum simulated temperature ranging from $T=0.42$ to 0.425 depending on $\varepsilon_{pd}$. $T_{g,d}$ is defined by $\tau_{Rd}(T_{g,d})=10^{14}\tau$, and $m_{f,d}$ is evaluated at the corresponding $T_{g,d}$. Horizontal lines at unity denote the polymer reference values. Error bars reflect propagation of the fitted $\tau_0$ uncertainty.}
\label{figS10}
\end{figure}

\clearpage
\begin{figure}
    \centering
    \includegraphics[width=1\linewidth]{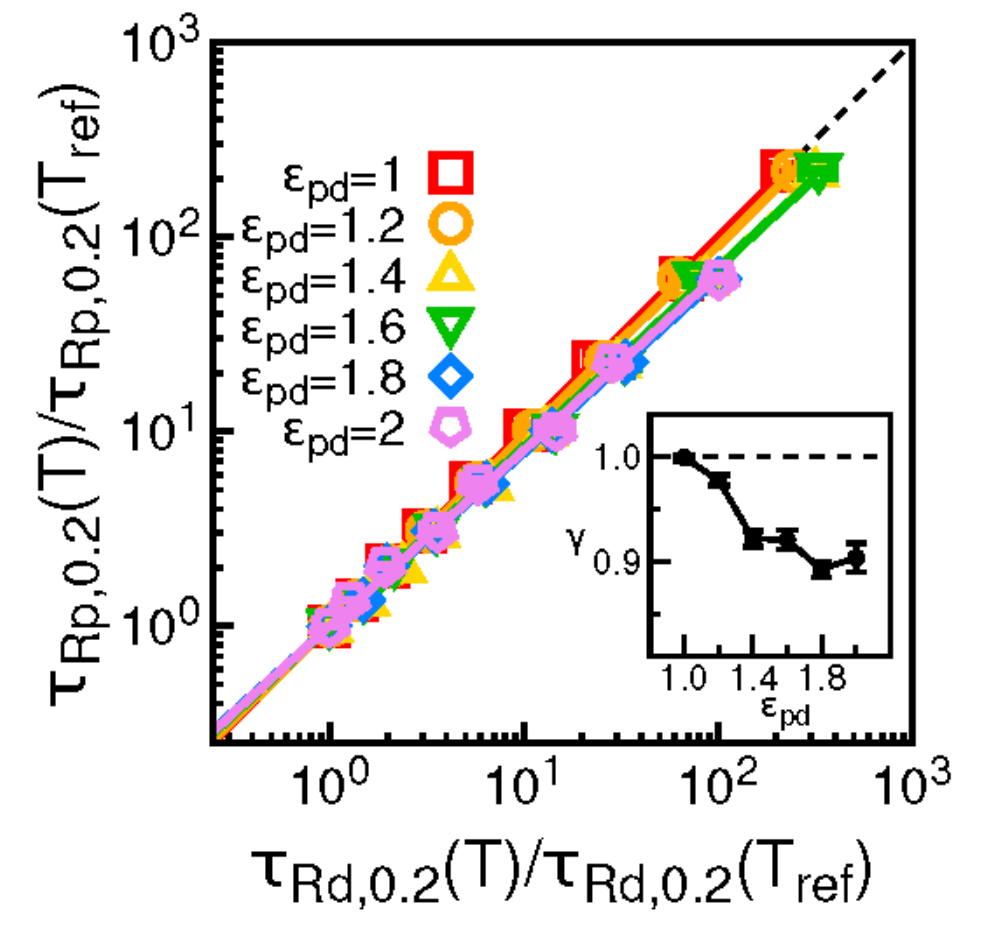}
\caption{Probe--host relaxation scaling using operationally defined relaxation times. 
The operational probe rotational relaxation time, $\tau_{Rd,0.2}$, and polymer bond-orientational relaxation time, $\tau_{Rp,0.2}$, are defined as the times at which the corresponding correlation functions decay to 0.2. Relative polymer relaxation, $\tau_{Rp,0.2}(T)/\tau_{Rp,0.2}(T_{\mathrm{ref}})$, is plotted against the corresponding relative probe relaxation, $\tau_{Rd,0.2}(T)/\tau_{Rd,0.2}(T_{\mathrm{ref}})$, for $\varepsilon_{pd}=1\varepsilon$--$2\varepsilon$, with $T_{\mathrm{ref}}=0.5$. Solid lines denote fits to the fractional scaling relation
$\tau_{Rp,0.2}(T)/\tau_{Rp,0.2}(T_{\mathrm{ref}})
=[\tau_{Rd,0.2}(T)/\tau_{Rd,0.2}(T_{\mathrm{ref}})]^{\gamma}$.
The diagonal dashed line denotes one-to-one scaling. The inset shows the corresponding scaling exponent $\gamma$ as a function of $\varepsilon_{pd}$; the horizontal dashed line denotes $\gamma=1$.}    
\label{figS11}
\end{figure}

\clearpage
\begin{figure}
    \centering
    \includegraphics[width=1\linewidth]{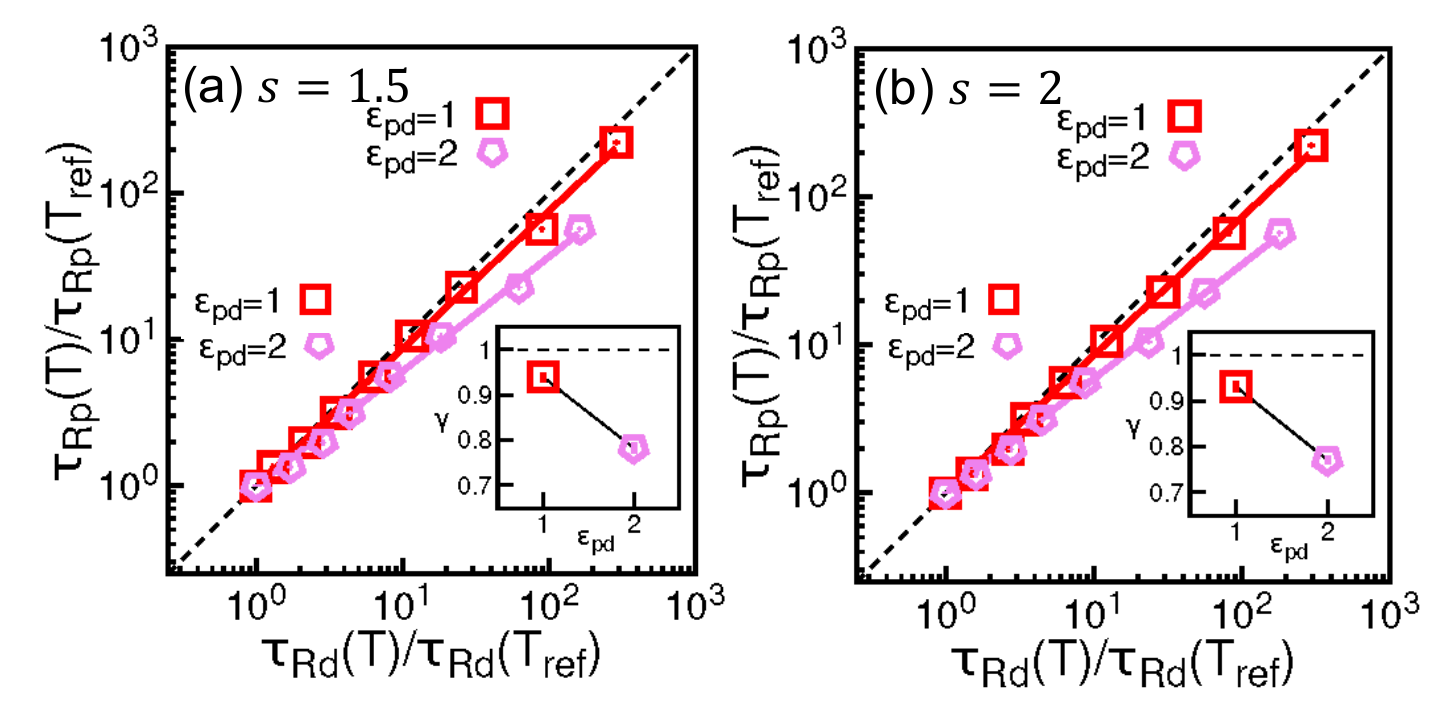}
    \caption{Probe--host relaxation scaling for larger dimer probes. 
Relative polymer bond-orientational relaxation time, $\tau_{Rp}(T)/\tau_{Rp}(T_{\mathrm{ref}})$, is plotted against the corresponding relative probe rotational relaxation time, $\tau_{Rd}(T)/\tau_{Rd}(T_{\mathrm{ref}})$, for dimer probes with (a) $s=1.5$ and (b) $s=2$, where $s$ denotes the LJ diameter of each probe bead relative to the polymer-monomer diameter, $\sigma_{dd}=s\sigma$. The probe--polymer interaction diameter is changed consistently using the same mixing rule as in the original model, $\sigma_{pd}=(\sigma_{pp}+\sigma_{dd})/2$, with $\sigma_{pp}=\sigma$. Results are shown for $\varepsilon_{pd}=1\varepsilon$ and $2\varepsilon$, and $T_{\mathrm{ref}}=0.5$ is used for normalization. Solid lines denote fits to the fractional scaling relation
$\tau_{Rp}(T)/\tau_{Rp}(T_{\mathrm{ref}})
=[\tau_{Rd}(T)/\tau_{Rd}(T_{\mathrm{ref}})]^{\gamma}$.
The diagonal dashed line denotes one-to-one scaling. Insets show the corresponding scaling exponent $\gamma$ for each interaction strength; the horizontal dashed line denotes $\gamma=1$.}
    \label{figS12}
\end{figure}

\clearpage
\begin{figure}
    \centering
    \includegraphics[width=1\linewidth]{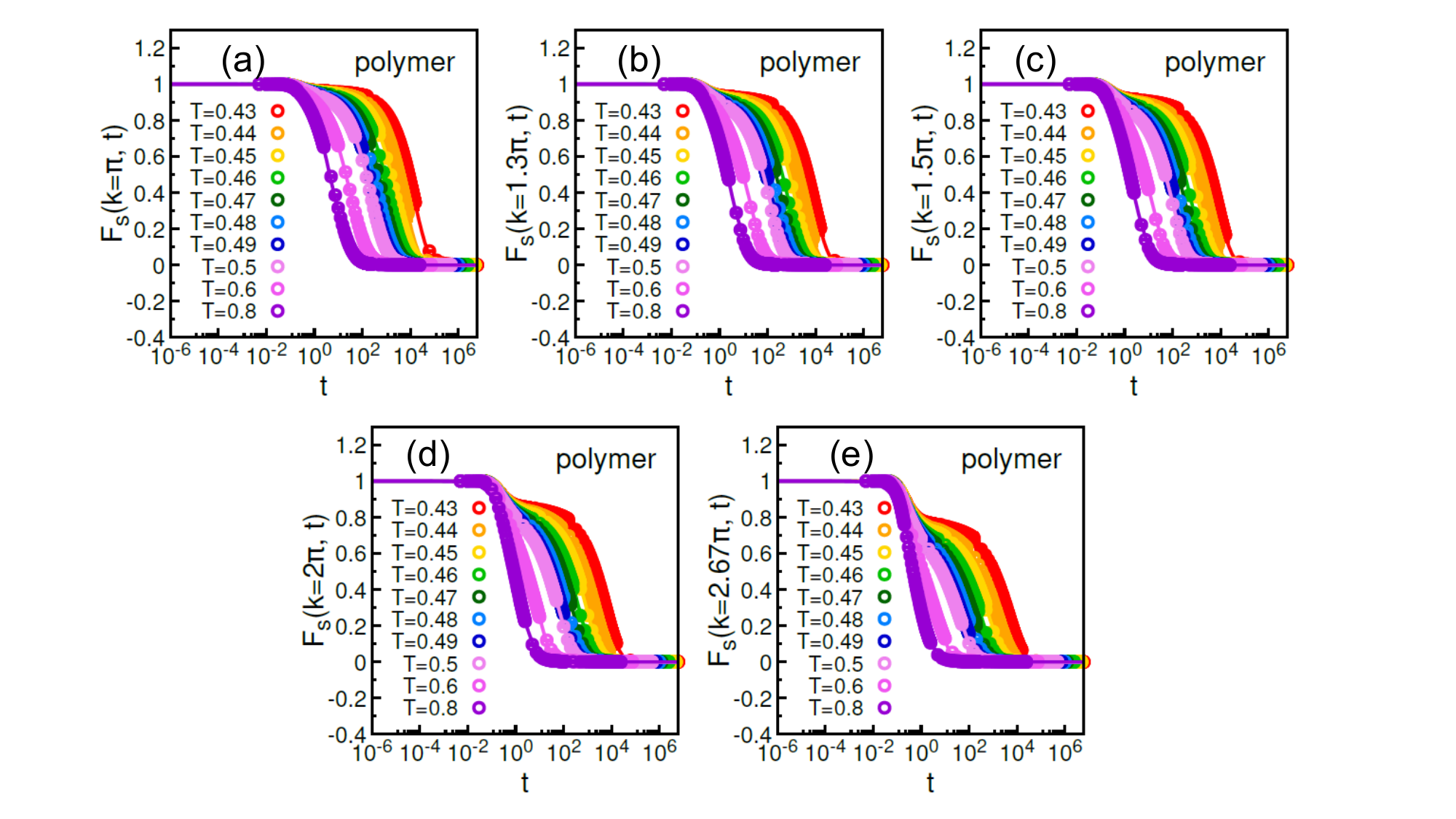}
    \caption{Wavevector-dependent structural relaxation of the bulk polymer. Self-intermediate scattering functions ($F_s(k,t)$) of the bulk polymer are shown at different temperatures for (a) $k=\pi/\sigma$, (b) $k=1.3\pi/\sigma$, (c) $k=1.5\pi/\sigma$, (d) $k=2\pi/\sigma$, and (e) $k=2.67\pi/\sigma$. Markers represent simulation results, and solid lines denote fits using a combination of exponential and stretched exponential functions.
}
    \label{figS13}
\end{figure}

\clearpage

\begin{figure}
    \centering
    \includegraphics[width=1\linewidth]{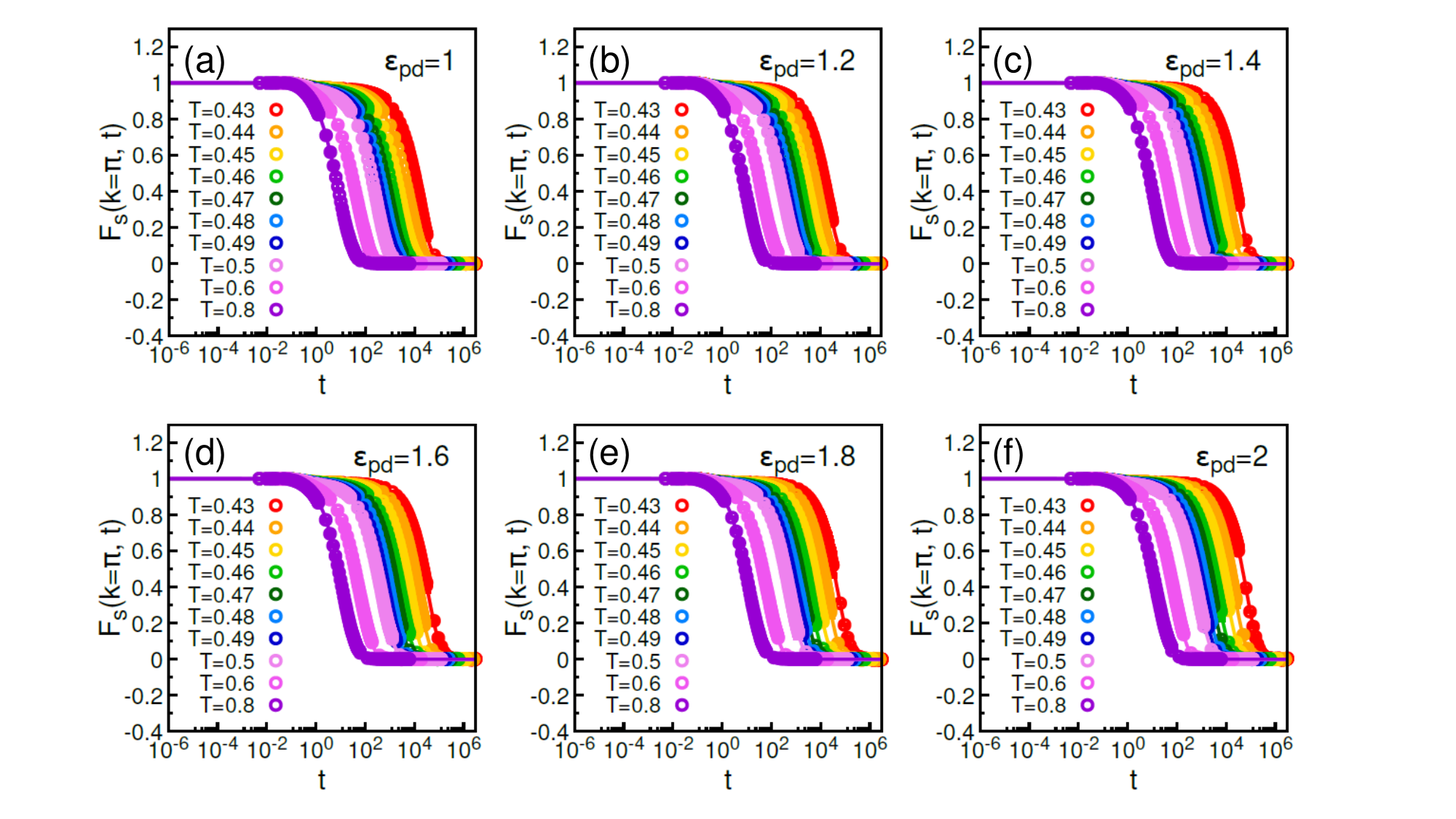}
    \caption{Probe dimer structural relaxation at $k=\pi/\sigma$. Self-intermediate scattering functions ($F_s(k,t)$) of the probe dimers at $k=\pi/\sigma$ are shown at different temperatures for (a) $\varepsilon_{pd}=1$, (b) $\varepsilon_{pd}=1.2$, (c) $\varepsilon_{pd}=1.4$, (d) $\varepsilon_{pd}=1.6$, (e) $\varepsilon_{pd}=1.8$, and (f) $\varepsilon_{pd}=2$. Markers represent simulation results, and solid lines denote fits using a combination of exponential and stretched exponential functions.
}
    \label{figS14}
\end{figure}

\clearpage
\begin{figure}
    \centering
    \includegraphics[width=1\linewidth]{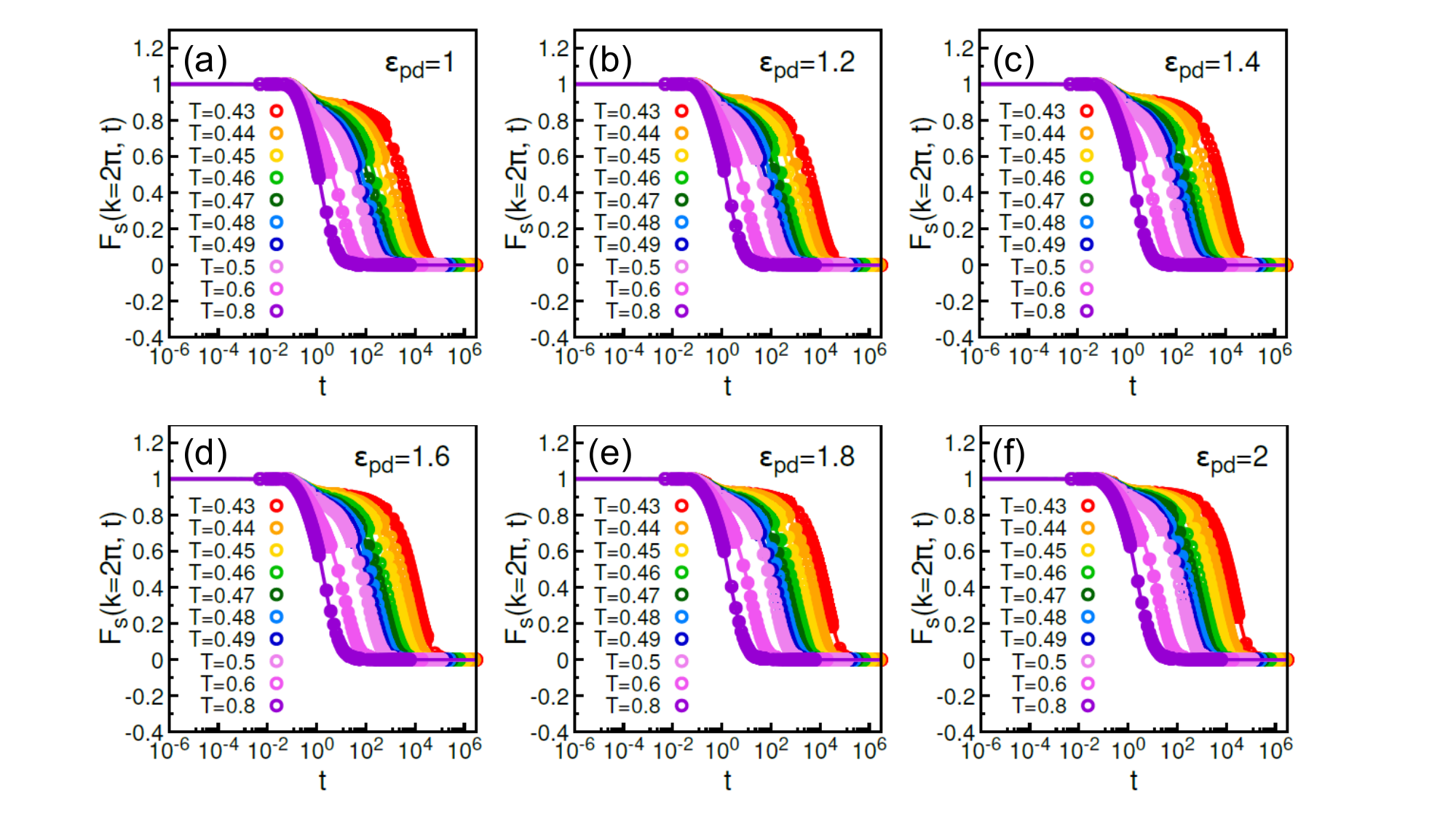}
    \caption{Probe dimer structural relaxation at $k=2\pi/\sigma$. Self-intermediate scattering functions ($F_s(k,t)$) of the probe dimers at $k=2\pi/\sigma$ are shown at different temperatures for (a) $\varepsilon_{pd}=1$, (b) $\varepsilon_{pd}=1.2$, (c) $\varepsilon_{pd}=1.4$, (d) $\varepsilon_{pd}=1.6$, (e) $\varepsilon_{pd}=1.8$, and (f) $\varepsilon_{pd}=2$. Markers represent simulation results, and solid lines denote fits using a combination of exponential and stretched exponential functions.
}
    \label{figS15}
\end{figure}

\clearpage
\begin{figure}
    \centering
    \includegraphics[width=1\linewidth]{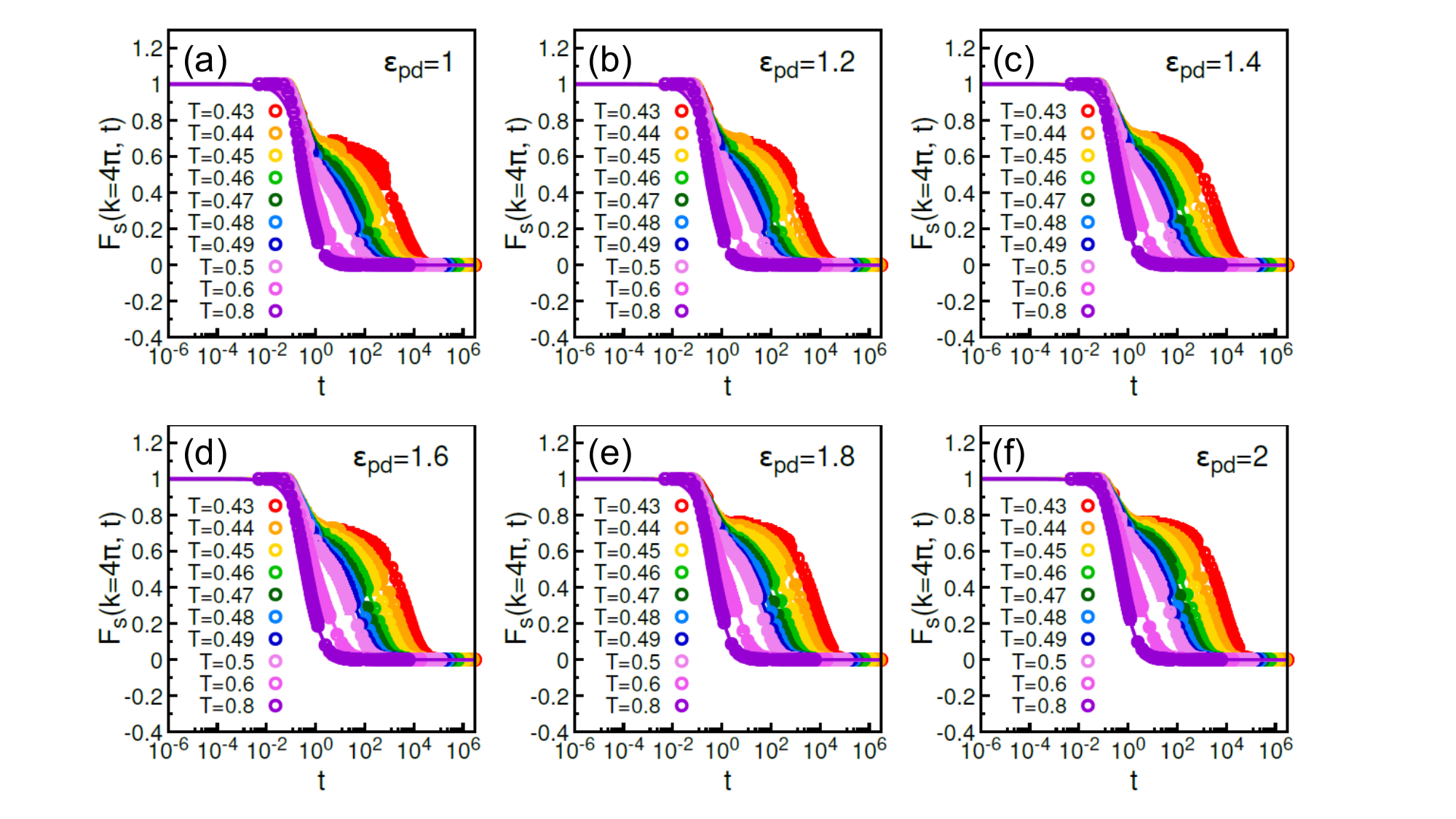}
    \caption{Probe dimer structural relaxation at $k=4\pi/\sigma$. Self-intermediate scattering functions ($F_s(k,t)$) of the probe dimers at $k=4\pi/\sigma$ are shown at different temperatures for (a) $\varepsilon_{pd}=1$, (b) $\varepsilon_{pd}=1.2$, (c) $\varepsilon_{pd}=1.4$, (d) $\varepsilon_{pd}=1.6$, (e) $\varepsilon_{pd}=1.8$, and (f) $\varepsilon_{pd}=2$. Markers represent simulation results, and solid lines denote fits using a combination of exponential and stretched exponential functions.}
    \label{figS16}
\end{figure}

\clearpage
\begin{figure}
    \centering
    \includegraphics[width=1\linewidth]{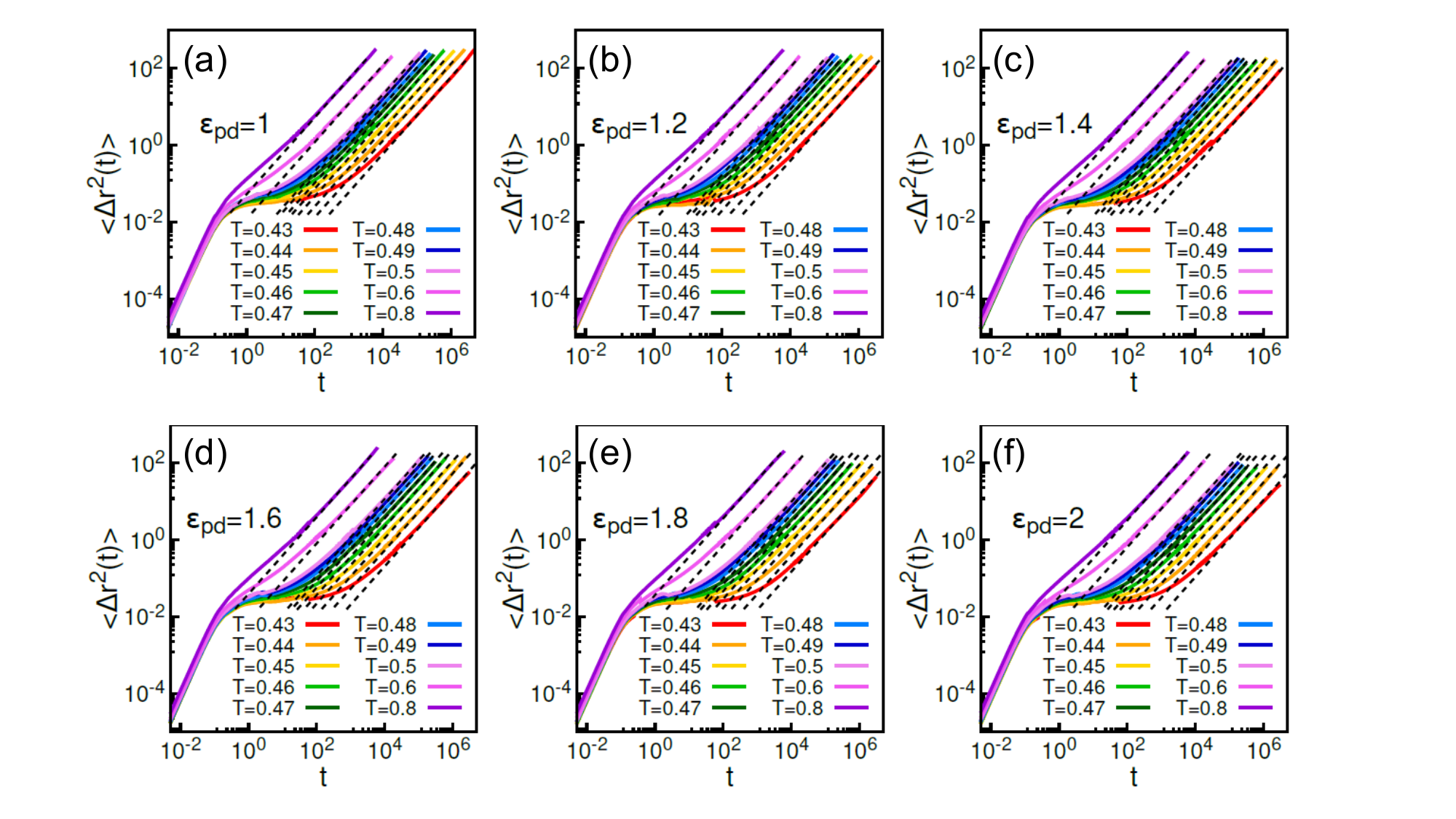}
    \caption{Mean-squared displacement of probe dimers. The mean-squared displacement $\langle \Delta r^2(t)\rangle$ of the probe dimer centers of mass is shown at different temperatures for (a) $\varepsilon_{pd}=1$, (b) $\varepsilon_{pd}=1.2$, (c) $\varepsilon_{pd}=1.4$, (d) $\varepsilon_{pd}=1.6$, (e) $\varepsilon_{pd}=1.8$, and (f) $\varepsilon_{pd}=2$. Dashed lines indicate fits to the long-time diffusive regime, $\langle \Delta r^2(t)\rangle = 6D_T t$.
}
    \label{figS17}
\end{figure}

\clearpage
\begin{figure}
    \centering
    \includegraphics[width=1\linewidth]{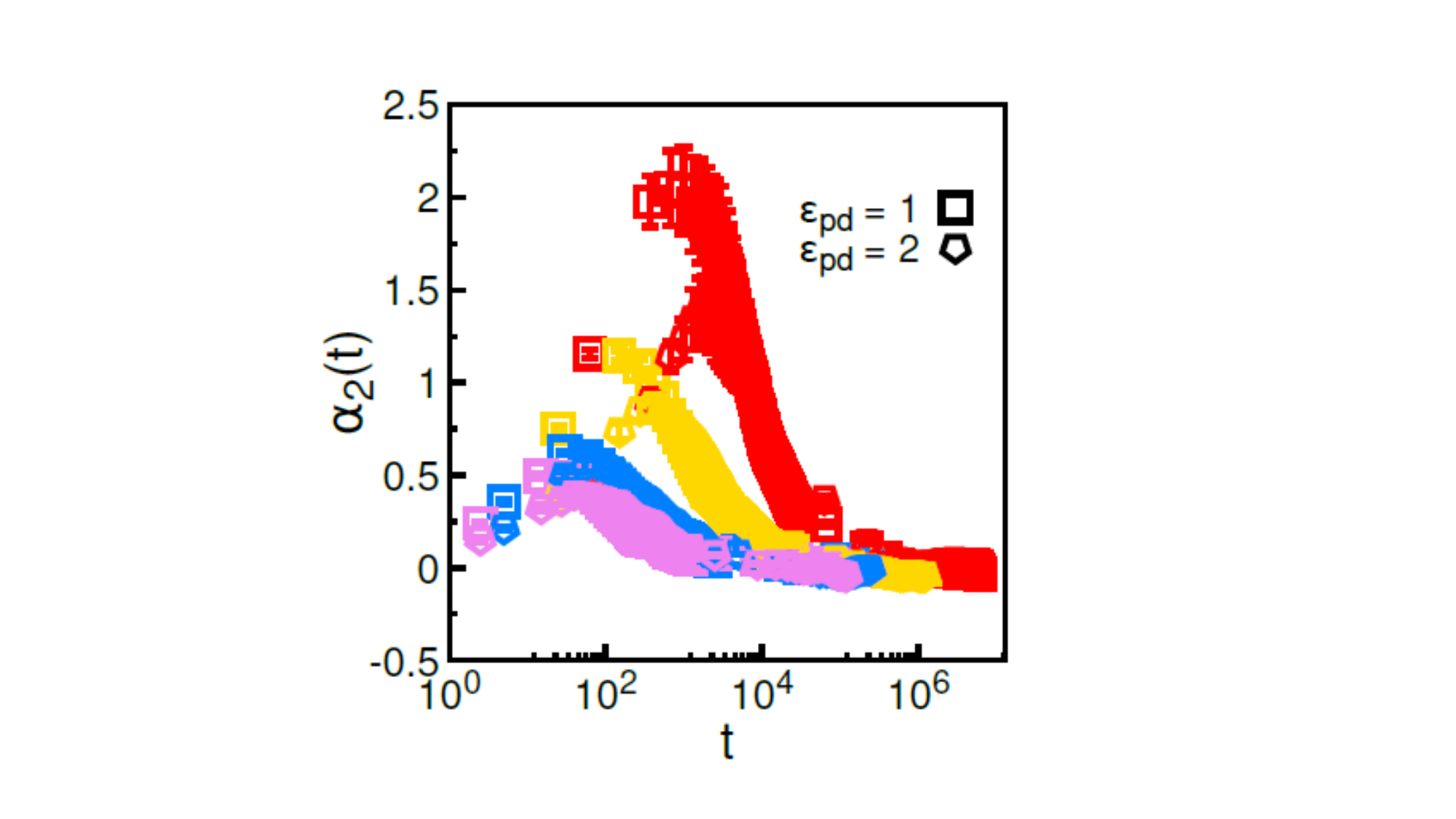}
    \caption{Non-Gaussian parameter of probe dimers with error bars. Time evolution of the non-Gaussian parameter $\alpha_2(t)$ is shown for weakly interacting probe dimers ($\varepsilon_{pd}=1$) and strongly interacting probe dimers ($\varepsilon_{pd}=2$) at different temperatures. 
}
    \label{figS18}
\end{figure}

\clearpage
\begin{figure}
    \centering
    \includegraphics[width=1\linewidth]{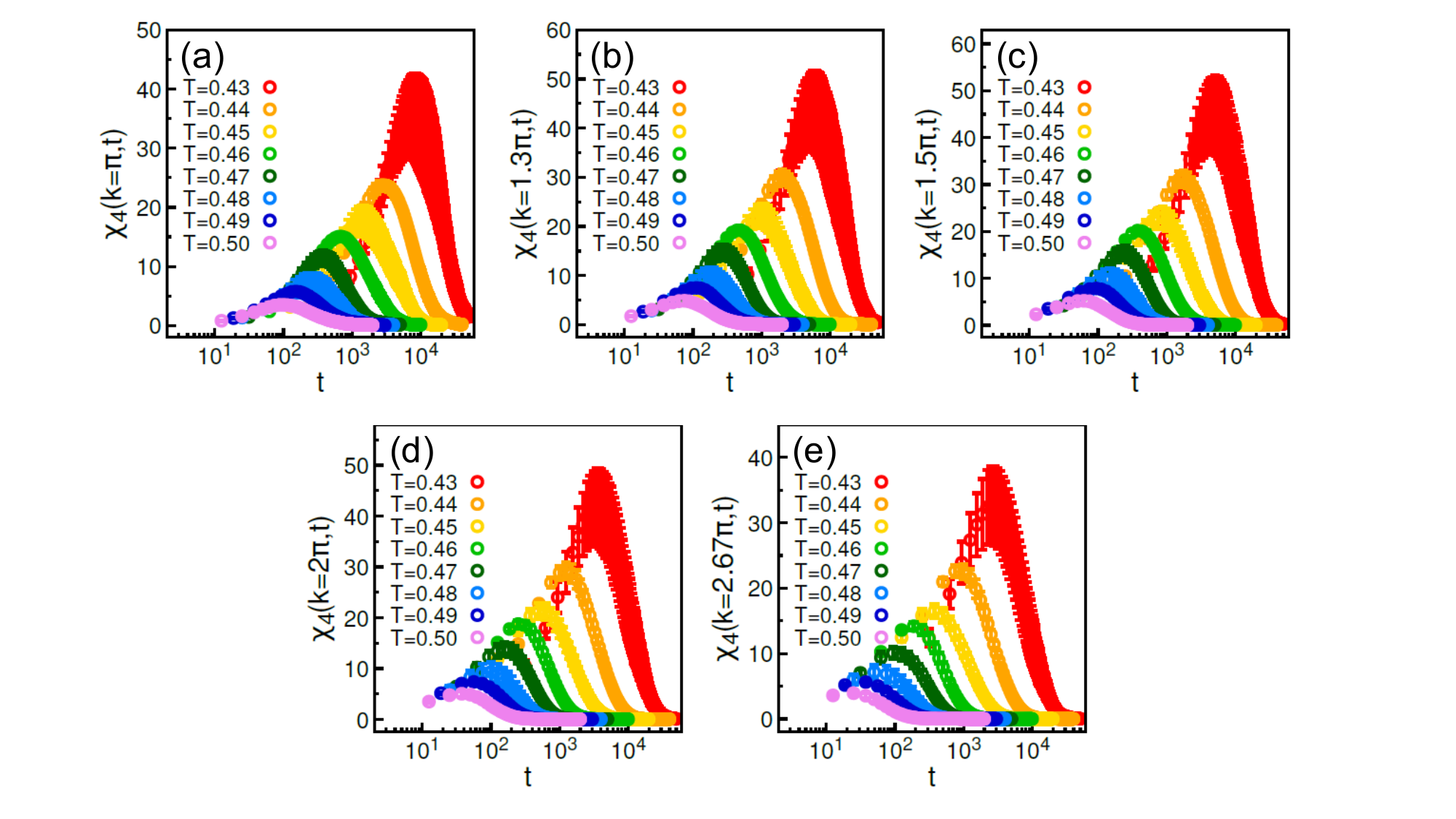}
    \caption{Wavevector-dependent four-point dynamic susceptibility of the bulk polymer. The time evolution of the $k$-dependent four-point dynamic susceptibility $\chi_4(k,t)$ is shown at different temperatures for (a) $k=\pi/\sigma$, (b) $k=1.3\pi/\sigma$, (c) $k=1.5\pi/\sigma$, (d) $k=2\pi/\sigma$, and (e) $k=2.67\pi/\sigma$.
}
    \label{figS19}
\end{figure}

\clearpage
\begin{figure}
    \centering
    \includegraphics[width=1\linewidth]{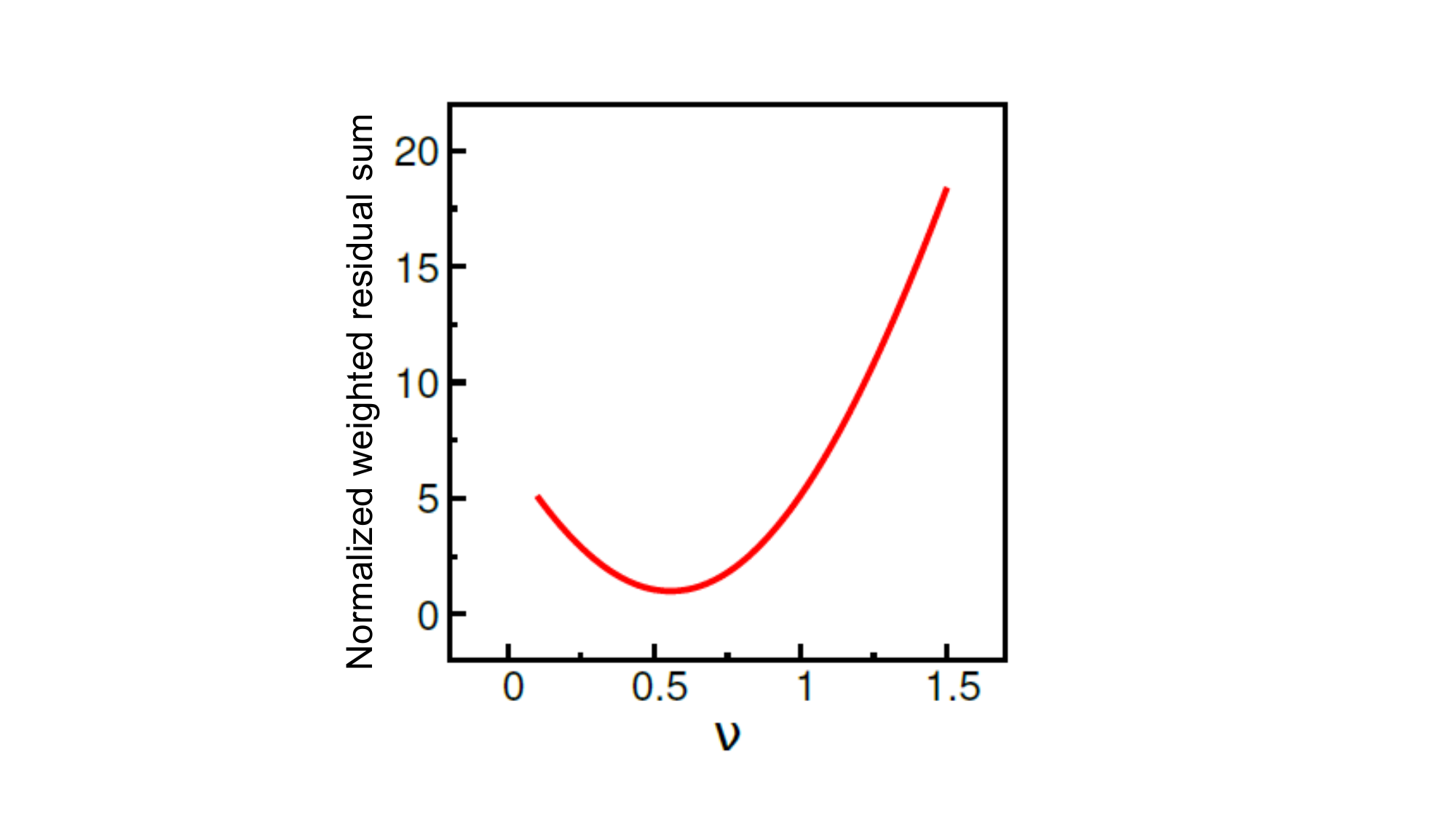}
    \caption{Determination of the optimal shared scaling exponent $\nu$. For each trial value of $\nu$, the transformed relaxation time data were fitted using weighted linear regression, and the resulting weighted residual sum was normalized by its minimum value. The minimum determines the optimal exponent $\nu=0.556$.
}
    \label{figS20}
\end{figure}

\clearpage
\begin{table}[htbp]
\centering
\caption{Fitted values of $\Delta\mu_k$ and $\Delta\mu_{\varepsilon}$ used in the dynamic heterogeneity scaling analysis. For the fixed exponent $\nu=0.556$, $\Delta\mu_k$ and $\Delta\mu_{\varepsilon}$ were obtained from weighted linear fits of the bulk polymer and probe dimer relaxation data, respectively. Each $\varepsilon_{pd}$ corresponds to the probe interaction strength assigned to the wavevector $k$.}
\label{tabS1}
\begin{tabular}{|c|c|c|c|}
\toprule
$k$ & $\Delta\mu_k$ & $\varepsilon_{pd}$ & $\Delta\mu_{\varepsilon}$ \\
\midrule
$\pi/\sigma$       & 0.334 & 1.0 & 0.337 \\
$1.3\pi/\sigma$    & 0.321 & 1.2 & 0.320 \\
$1.5\pi/\sigma$    & 0.322 & 1.4 & 0.329 \\
$2\pi/\sigma$      & 0.356 & 1.6 & 0.357 \\
$2.67\pi/\sigma$   & 0.431 & 2.0 & 0.423 \\
\bottomrule
\end{tabular}
\end{table}

\end{document}